\documentclass[sigconf, screen=true]{acmart}
\AtBeginDocument{%
  }

\setcopyright{acmcopyright}
\copyrightyear{2024}
\acmYear{2024}
\acmDOI{XXXXXXX.XXXXXXX}
\acmISBN{978-1-4503-XXXX-X/18/06}

\usepackage[utf8]{inputenc}
\usepackage{color}
\usepackage{tikz}
\usepackage{amsmath}

\usepackage[ruled,linesnumbered]{algorithm2e}
\usepackage{flexisym}
\usepackage{breqn}
\usepackage{cleveref}
\usepackage{booktabs}
\usepackage[most]{tcolorbox}

\usepackage{xspace}
\usepackage{pifont}
\usepackage{caption}
\usepackage{subcaption}
\usepackage{makecell}
\usepackage[flushleft]{threeparttable}
\usepackage{multirow}
\usepackage{multicol}
\usepackage{xcolor, colortbl}
\usepackage{float} 
\usepackage{bbding}
\usepackage{enumitem}
\usepackage{setspace}
\usepackage{xurl}
\usepackage{array}

\usepackage[adversary,advantage, operators,sets,events, keys, probability]{cryptocode}
\usepackage[colorinlistoftodos,prependcaption]{todonotes}

\crefformat{section}{\S#2#1#3} 
\crefformat{subsection}{\S#2#1#3}
\crefformat{subsubsection}{\S#2#1#3}

\newtcolorbox{mybox}[2][]{colback=white, colframe=black, coltitle=black, fonttitle=\bfseries, boxrule=1.0pt,
colbacktitle=white, enhanced, attach boxed title to top right={xshift=-5mm, yshift=-2mm}, after skip=0pt, title={#2},#1}

\tcbset{
	thmbox/.style={
        enhanced,
        breakable,
        sharp corners=all,
        fonttitle=\bfseries\normalsize,
        fontupper=\normalsize\itshape,
         top=0mm,
         bottom=0mm,
         right=0mm,
        colback=white,
        colframe=white,
        colbacktitle=white,
        coltitle=black,
        attach boxed title to top left,
        boxed title style={empty, size=minimal, bottom=1.5mm},
        overlay unbroken ={
            \draw (title.south west)--(title.south east);
            \draw ([xshift=3.5mm]frame.north west)|-%
                  (frame.south east)--(frame.north east);},
        overlay first={
            \draw (title.south west)--(title.south east); 
            \draw ([xshift=3.5mm]frame.north west)--([xshift=3.5mm]frame.south west);
            \draw (frame.north east)--(frame.south east);},
        overlay middle={
            \draw ([xshift=3.5mm]frame.north west)--([xshift=3.5mm]frame.south west);
            \draw (frame.north east)--(frame.south east);},
        overlay last={
            \draw ([xshift=3.5mm]frame.north west)|-%
                  (frame.south east)--(frame.north east);},
        },
    LQ/.style={thmbox, 
        overlay unbroken ={
            \draw (title.south west)--(title.south east);
            \draw ([xshift=3.5mm]frame.north west)|-([xshift=15mm]frame.south west);
            \node[anchor=east] at (frame.south east) {};},
        overlay first={
            \draw (title.south west)--(title.south east); 
            \draw ([xshift=3.5mm]frame.north west)--([xshift=3.5mm]frame.south west);},
        overlay middle={
            \draw ([xshift=3.5mm]frame.north west)--([xshift=3.5mm]frame.south west);},
        overlay last={
            \draw ([xshift=3.5mm]frame.north west)|-([xshift=15mm]frame.south west);
            \node[anchor=east] at (frame.south east) {};},
    }, 
}

\newtcbtheorem[]{obsv}{}{thmbox,LQ}{theo}

\newcommand{\Aa}{\mathcal{A}}
\newcommand{\Bb}{\mathcal{B}}
\newcommand{\Vv}{\mathcal{V}}
\newcommand{\Pp}{\mathcal{P}}
\newcommand{\Oo}{\mathcal{O}}
\newcommand{\Dd}{\mathcal{D}}
\newcommand{\Ll}{\mathcal{L}}
\newcommand{\acore}{{\scshape Armored Core}}
\newcommand{\puf}{\textit{puf}}
\newcommand{\adver}{\mathcal{A}\mathit{dv}}
\newcommand{\lgr}{\mathcal{L}\mathit{gr}}

\newtheorem{theorem}{Theorem}

\newtheorem{definition}{Definition}

\newcommand{\PreserveBackslash}[1]{\let\temp=\\#1\let\\=\temp}
\newcolumntype{C}[1]{>{\PreserveBackslash\centering}p{#1}}
\newcolumntype{R}[1]{>{\PreserveBackslash\raggedleft}p{#1}}
\newcolumntype{L}[1]{>{\PreserveBackslash\raggedright}p{#1}}

\makeatletter
\DeclareRobustCommand\onedot{\futurelet\@let@token\@onedot}
\def\@onedot{\ifx\@let@token.\else.\null\fi\xspace}

\def\eg{\emph{e.g}\onedot} 

\def\ie{\emph{i.e}\onedot}

\def\etal{\emph{et al}\onedot}

\definecolor{cardinal}{rgb}{0.77, 0.12, 0.23}
\definecolor{ceruleanblue}{rgb}{0.16, 0.32, 0.75}
\definecolor{lavenderblue}{rgb}{0.8, 0.8, 1.0}
\definecolor{janusblue}{RGB}{96, 136, 172}
\definecolor{janusbrown}{RGB}{132, 78, 1}
\definecolor{acoreblue}{HTML}{889FC8}
\definecolor{commentblue}{HTML}{304486}

\newcommand*{\priority}[1]{\begin{tikzpicture}[scale=0.13]%
		\draw (0,0) circle (1);
		\fill[fill opacity=1,fill=acoreblue] (0,0) -- (90:1) arc (90:90-#1*3.6:1) -- cycle;
\end{tikzpicture}}

\newcommand{\nosemic}{\renewcommand{\@endalgocfline}{\relax}}
\let\oldnl\nl
\newcommand{\nonl}{\renewcommand{\nl}{\let\nl\oldnl}}

\SetCommentSty{mycommfont}
\SetArgSty{textnormal}
\SetKwInOut{Input}{Input}
\SetKwInOut{Output}{Output}
\SetKw{False}{false}
\SetKw{True}{true}
\SetKw{Valid}{valid}
\SetKw{Invalid}{invalid}

\newcommand{\compfull}{\priority{100}}
\newcommand{\comppart}{\priority{50}}
\newcommand{\compnone}{\priority{0}}

\begin{document}


\title{Now~Let's~Make~It~Physical:~Enabling~Physically~Trusted Certificate~Issuance~for~Keyless~Security~in~CAs}
%


\author{Xiaolin Zhang}
\affiliation{%
  \institution{Shanghai Jiao Tong University}
  \city{Shanghai}
  \country{China}
}

\author{Chenghao Chen}
\affiliation{%
  \institution{Shanghai University}
  \city{Shanghai}
  \country{China}
}

\author{Kailun Qin}
\affiliation{%
  \institution{Shanghai Jiao Tong University}
  \city{Shanghai}
  \country{China}
}

\author{Yuxuan Wang}
\affiliation{%
  \institution{Shanghai Jiao Tong University}
  \city{Shanghai}
  \country{China}
}

\author{Shipei Qu}
\affiliation{%
  \institution{Shanghai Jiao Tong University}
  \city{Shanghai}
  \country{China}
}

\author{Tengfei Wang}
\affiliation{%
  \institution{Shanghai Jiao Tong University}
  \city{Shanghai}
  \country{China}
}

\author{Chi Zhang}
\authornotemark[1]
\affiliation{%
  \institution{Shanghai Jiao Tong University}
  \city{Shanghai}
  \country{China}
}

\author{Dawu Gu}
\authornote{Corresponding authors}
\affiliation{%
  \institution{Shanghai Jiao Tong University}
  \city{Shanghai}
  \country{China}
}








\renewcommand{\shortauthors}{Trovato et al.}

\begin{abstract}
  The signing key protection of Certificate Authorities (CAs) remains a critical challenge in PKI. Traditional approaches struggle to eliminate the risk of key exposure due to those (un)intentional human errors. This long-standing dilemma motivates us to propose \acore, a novel PKI security extension using the trusted binding of Physically Unclonable Function (PUF) for CAs. PUFs leverage manufacturing variations to generate unique and random responses. Combining with XOR and hash, they can make key exposure impossible for CAs through keyless certificate issuance. 
  

  
  In \acore, we design a set of PUF-based X.509v3 certificate functions for CAs to generate physically trusted ``signatures'' without using a digital key. Moreover, we introduce a novel PUF transparency mechanism to effectively monitor the PUF operations in CAs. We integrate \acore\ into real-world PKI systems including Let's Encrypt Pebble and Certbot. We also provide a PUF-embedded hardware prototype. The evaluation results show that \acore\ can achieve keyless certificate issuance while improving the computation performance by 4.9\%$\sim$73.7\%. It only incurs small communication and storage overhead ($<4\%$).
\end{abstract}




\maketitle

\section{Introduction}\label{sec:introduction}







Public Key Infrastructure (PKI) heavily relies on Certificate Authorities (CAs) that sign the digital certificates, providing the binding of a public key to an entity. Protecting the long-term signing keys of CAs is crucial in PKI. However, keeping these keys private is quite challenging in practice. A single leaked key is sufficient to destroy the trust on a PKI vendor. In 2011, the attacker stole the signing key of DigiNotar \cite{diginotarreport} and issued 531 rogue certificates for famous domains like \texttt{google.com}. Trustico leaked signing keys \cite{trusticokeyleak} via email in 2018, resulting in 23,000 certificates being compromised. The code-signing keys of Nvidia \cite{nvdiakeyleak}, Samsung \cite{samsungkeyleak}, and MSI \cite{msikeyleak} were also exposed in recent years. Such incidents can still happen even today \cite{RecentCAKeyCompromise} with better security protections.


\begin{figure*}
    \centering
    \includegraphics[width=0.95\linewidth]{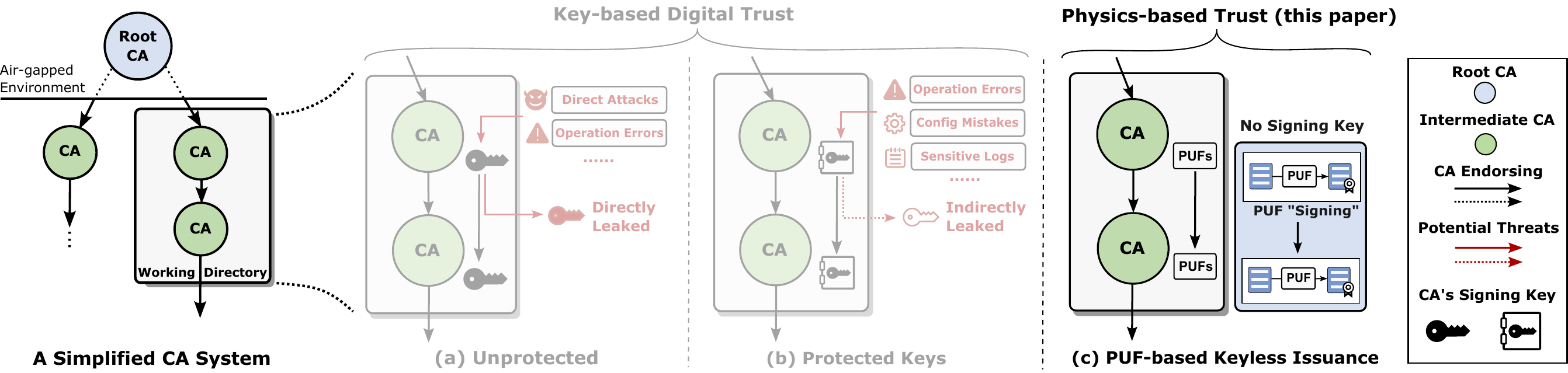}
    \caption{Comparison of existing approaches on CA certificate issuance in PKI.}\label{fig:pki_design_compare}
\end{figure*}


%




The protection of CAs' signing keys remains a persistent concern in network systems, particularly for those intermediate CAs interacting with the outside world. Current protections usually fall into two ways: \ding{172} \textit{split the key} or \ding{173} \textit{seal the key}. For \ding{172}, threshold signature \cite{KubilayKM21KORGAN, RPKIThreshold, ThresholdCCS20DKG, ThresholdCCS23, DalskovOKSS20} can split the signing key into multiple shares. However, the operations cross multiple CAs do not fit into current PKI architecture. On the other hand (\ding{173}), PKI vendors prefer to adopt the secure hardware like Hardware Security Modules (HSM) \cite{cloudhsm} and Trusted Execution Environments (TEE) to isolate the CA's signing key, which are effective against most attack vectors. However, in many real incidents above, the key was exposed more due to non-technical errors. Sloppy configurations \cite{CVE-2015-7328,CVE-2018-17612,CVE-2020-28053}, careless privileges \cite{CVE-2022-23716,CVE-2017-6339}, undeleted keys in codebase \cite{gitsecretTrustCom, Dockerhubsecret} and other human mistakes could all cause to the exposure of the CA signing key. 


Although approaches like HSMs are quite secure in most cases, but those operational errors could render them useless. In this ``\textit{expose-protect-expose again}" dilemma, we capture two critical observations (\cref{sec:2_motivation}): \ding{172} As long as a digital signing key exists, there is a possibility of it being exposed in some ways. \ding{173} One single exposure can cause a disaster. Once attackers obtains that key string, they can issue certificates with almost no limits. They motivate us to consider a bold research question: \textbf{Is it possible to issue certificates without an explicit digital key in CAs?}

Our core insight in answering this question is to turn \textit{key-based digital trust} into \textit{keyless physical trust} as shown in Figure \ref{fig:pki_design_compare}, thereby eliminating the possibility of the digital leakage for keys. We notice that Physically Unclonable Function (PUF), a novel hardware security primitive, can leverage random manufacturing variations to map input challenges to unique responses. The \textit{independent trust} and \textit{physical authenticity} it provides (\cref{sec:3_puf_advantage}) exactly match our envision. PUF can establish trust itself without relying on a sealed key.

PUF has been widely validated by many semiconductor giants including Xilinx \cite{xilinxpuf}, NXP \cite{nxplpcpuf}, Samsung \cite{SamsungPUF} and Nvidia \cite{NvidiaPUF}. It is often treated as a key generator \cite{PUFSSL2018,PUFPKIIoT2022,intrinsicidpuf,NvidiaPUF} to avoid the secure storage. This usage, however, does not eliminate the digital footprint when the key is explicitly used, and the system still cannot afford any leakage of that key string. In academia, many security schemes \cite{chatterjeeBuildingPufBased2019,chaterjee3PAARivateUF2021,xiaolinSPEAR,ghaeiniPAttPhysicsbasedAttestation,zhang2024teamwork,PUFUCBrzuskaFSK11} regard PUF as a stable cryptographic oracle to bind the entity with its Challenge-Response Pairs (CRPs). So far in PKI, except for the key generation \cite{PUFPKIIoT2022}, the exploration of PUF's security advantages is insufficient.

In this work, we treat PUF as a physical signing primitive (Figure \ref{fig:pki_design_compare}(c)) and propose \acore\footnote{Armored Core is a popular game series developed by FromSoftware.}, a novel PUF-based PKI security extension. At a high level, the certificate to be signed is directly fed into PUF to generate the ``signature'' response without a sealed signing key in CAs. The workflow of \acore\ is similar to the traditional PKI and the keys are still needed in other PKI entities. However, this unconventional design for PKI would bring three challenges. First, using PUF for certificate ``signing'' and issuance can cause significant compatibility issues with the original certificates. Second, user agents cannot interact with the CA (and its PUFs), so they have to store enough correct CRPs in advance for verification. This further leads to the third challenge: The invocation of PUFs of CAs is not publicly auditable. How to mitigate the abuse of PUFs have not been discussed in prior studies.

To address these challenges, we first develop a cryptographic-style theoretical abstraction of PUF in PKI/CA. It decouples CA's functions with the PUF hardware so that \acore\ can be PUF-agnostic. Any actual PUF designs that conform to this abstraction can be used. Based on it, we propose a PUF-based certificate chain construction (\cref{sec:4_puf_certificate_ops}) that is interoperable with standard certificates to deal with the first challenge. It only makes minimal modifications to X.509v3 fields with PUF-backed endorsements. For the second and third challenge, we offer a PUF transparency logging mechanism (\cref{sec:4_puf_transparency}) where the PUF invocation vector is defined to record the PUF calling behaviors. This mechanism can be deployed with Certificate Transparency (CT) to enable in-time CRP verification and mitigate potential PUF abuses.

We implement two open-sourced prototypes \cite{acoregithub} of \acore\ (\cref{sec:implementation}): a full functionality prototype integrated into real-world PKI software including Let's Encrypt Pebble CA \cite{pebble}, Certbot, and Google Trillian \cite{trillian} log server, and a hardware prototype of RISC-V CPU \cite{RISCVCVA6} with a built-in Interpose PUF (IPUF) \cite{InterposePUF19CHES} that can be installed in network devices. We examine the system security and present formal cryptographic proofs (\cref{sec:full_proof}). We also comprehensively evaluate our design (\cref{sec:evaluation}) from PUF performance, design comparison, microbenchmark, introduced overhead and deployment cost. The results show that it enhances security while surprisingly improves the computing efficiency by 4.9\%$\sim$73.7\%.


Compared with existing protections, \acore\ establishes security by intrinsic structures of PUF to internalize physical randomness, instead of by provisioning a sealed key in CAs. With it, various attacks will be in vain and operational errors will be nonfatal since there is no key to be leaked in CAs. Besides, even if attackers compromise the whole CA software, they must rely on the specific PUF hardware in that server to issue certificates. Therefore, \acore\ can greatly reduce the \textit{post-compromise} damage (discussed in \cref{sec:discussion}), which validates its design strategy.













In summary, our contributions are as follows: 

\begin{itemize}
    \item \textbf{New Research Question}: We raise and address a novel and risky research question. The solution to it can make various attacks ineffective and establish a more trusted PKI ecosystem with existing protection techniques.
    \item \textbf{New Security Design for PKI}: We propose \acore, the first PKI security extension allowing CAs to issue public key certificates by PUF-based physically trusted bindings, which effectively eliminates the risk of key exposure.
    \item \textbf{Practical Prototypes}: We integrate \acore\ into real-world PKI codebase and develop a PUF-embedded RISC-V CPU prototype, which lays the foundation for future deployment. We thoroughly evaluate it and the results confirm its efficiency and effectiveness in practice.
\end{itemize}

\section{Background \& Motivation}\label{sec:2_motivation}

\begin{figure*}
  \centering
  \includegraphics*[width=0.95\linewidth]{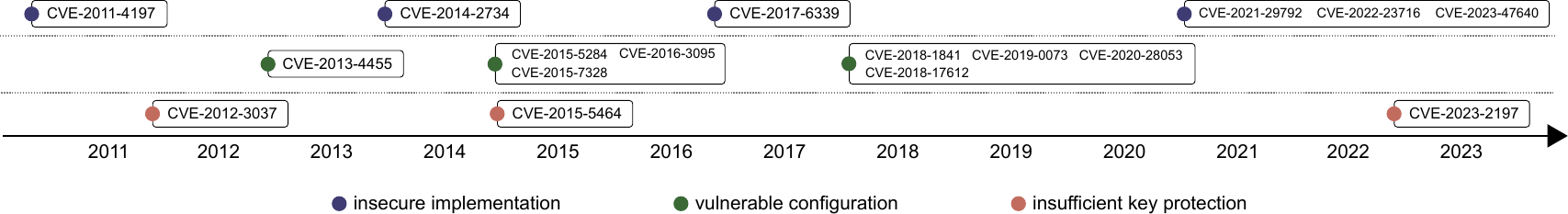}
  \caption{Technical vulnerabilities that cause the exposure of CAs' signing keys since 2010s. \ding{172} \textbf{Insecure implementation}: Flawed logic \cite{CVE-2011-4197,CVE-2017-6339,CVE-2023-47640} and uncensored log \cite{CVE-2021-29792,CVE-2022-23716}. \ding{173} \textbf{Vulnerable configuration} \cite{CVE-2015-7328,CVE-2018-17612,CVE-2020-28053}: The signing key file is erroneously configured to be publicly readable. \ding{174} \textbf{Insufficient key protection}: CAs store the keys without strong protections \cite{CVE-2012-3037}.}\label{fig:cve_summarize}
\end{figure*}

\subsection{PKI and Certificate Transparency}\label{sec:21_certificate_transparency}

PKI had been the \textit{de facto} standard technique for entity authentication. CAs endorse an entity through the cryptographic binding of a digital signature. They establish trust on the assurance that the private signing key has been securely kept in CA. To mitigate the certificate misissuance, Google introduced CT \cite{rfc9162}, which has been a necessary trusted component in modern PKIs.

CT involves semi-public distributed loggers to maintain append-only logs. These loggers are usually deployed by CA vendors\footnote{\url{https://certificate.transparency.dev/logs/}} in practice. CAs must submit the certificates to them to acquire the Signed Certificate Timestamps (SCTs) which serve as a cryptographic proof of the submission. Users can obtain an SCT in three ways \cite{rfc9162}. \ding{172} Certificate extension: The SCT is embedded in the certificate as an X.509 extension. \ding{173} OCSP stapling: It is included in a signed OCSP response from the domain server. \ding{174} TLS stapling: It is sent with the TLS handshake message. Typically, to maintain the consistent trust level, loggers can be configured to allow only permitted CAs to submit \cite{googleCTpolicy} but keep the query ports open.








\subsection{Signing Key Exposure of CA}\label{sec:22_key_exposure_CA}




The exposure of long-term signing key has been one of the worst incidents in PKI. A surprising fact is that they have occurred from time to time, even just recently \cite{RecentCAKeyCompromise}. There have been several disclosed incidents including DigiNotar in 2011 \cite{diginotarreport}, CNNIC in 2014 \cite{CNNICkeyleak}, Lenovo in 2015 \cite{CISAlenovo} and Trustico in 2018 \cite{trusticokeyleak}. In code-signing PKI, Bit9 \cite{Bit9keyleak}, D-link \cite{Dlinkkeyleak}, Nvidia \cite{nvdiakeyleak}, Samsung \cite{samsungkeyleak}, MediaTek \cite{mediatekkeyleak} and MSI \cite{msikeyleak} have all encountered the exposure of their signing keys. There could be more undisclosed incidents yet to be discovered. The causes of these incidents are diverse and not purely technical. Careless operators, wrong configurations and sloppy management can result in the exposure or theft of these keys \cite{Serrano2019PKIincident}.




On the other side, the technical vulnerabilities of CAs are rare to be seen but they do not disappear yet. We summarize and classify the CVEs that can cause the \textit{direct} exposure of CA's signing keys in Figure \ref{fig:cve_summarize}. Though many of these CVEs have been fixed and are not publicly exploitable, but this risk for PKI/CA has not diminished. Meanwhile, we observe that there is \textit{no} apparent relations between them. For example, \textsc{CVE-2015-7328} is caused by the Puppet Server's vulnerabilities; \textsc{CVE-2018-17612} involves writing the private key into a public \texttt{pem} file; \textsc{CVE-2020-28053} is due to a wrongly configured command on key exporting. This is possibly the reason why it is difficult to avoid all kinds of exposure in practice. In summary, we give the following observation.

\begin{obsv*}{Observation 2.1}{}\label{obs:2-1}

  As many incidents and CVEs have indicated, the exposure risk of signing keys of CA continues to persist. Even today, many non-technical reasons can all lead to the (in)direct key leakage.
\end{obsv*}






\subsection{Existing Mitigations and Limitations}\label{sec:existing_problem}




Signing key protection had been a clich\'e for PKI, but it seems never having an effective solution. To mitigate such critical risk, both academia and industry have developed several approaches.

\ding{172} \textbf{Split the key}: Decentralized PKI \cite{KubilayKM19CertLedger} operations can prevent the single point of failure, where threshold signature and blockchain are two common techniques \cite{KubilayKM21KORGAN, TooraniG21DecentralizedPKI}. However, decentralized PKIs require a complete overhaul of the architecture, which is incompatible and could bring much overhead \cite{SermpinisVKV21DeTRACT, KubilayKM19CertLedger}. 

\ding{173} \textbf{Seal the key}: The real-world PKI vendors like Let's Encrypt, WoSign and Digicert prefer to use secure hardware, \eg, HSM, to isolate the key and signing operation. It is effective against many attacks \cite{CVE-2015-7328, CVE-2017-6339} and has been widely used for years. Such protection is sufficient under most cases since the keys never leave the hardware. However, the leakage in many incidents was not caused by direct attacks on the protections themselves. It is a kind of \textit{indirect} exposure due to operational errors, wrong configurations \cite{CVE-2020-28053, CVE-2023-2197} and privileges \cite{CVE-2020-26155}, which renders well-thought-out defense useless. This problem is not because HSMs or TEEs are not secure enough, but simply that it is beyond what they should cover.

\noindent$\bullet$ \textbf{Dilemma.} Though the existing protections for signing keys have been highly secure against various attacks, the incidents of key compromise still happen sometimes. They can effectively reduce but not eliminate such risk. We argue that the fact that the CA's key is used digitally causes the possibility of its exposure, but ensuring its permanent security across all links is quite onerous. Besides, the current situation is more favorable for attackers. With just one exposed signing key, they can break that CA and generate certificates anywhere. This dilemma motivates us to offer this observation and further think: \textit{Can we not use a digital signing key in CA? How to ensure the authenticity after then? How can the trust be established?}

\begin{obsv*}{Observation 2.2}{}

  The existing approaches are technically secure but could be bypassed due to various human mistakes. The existence of a digital key in CAs makes it always possible for attackers to identify the potential leakage to recover the key string. 
\end{obsv*}




\section{PUF for CA: Build Trust Without Keys}\label{sec:3_puf_advantage}

\begin{figure}[htbp]
  \centering
  \begin{minipage}[c]{0.65\columnwidth}
    \centering
    \includegraphics[width=\textwidth]{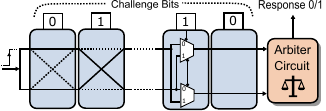}
		\subcaption{Basic arbiter PUF construction}
		\label{fig:puf_basic_a}
  \end{minipage}\qquad
  \begin{minipage}[c]{0.23\columnwidth}
    \centering
    \includegraphics[width=\textwidth]{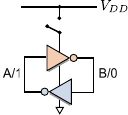}
    \subcaption{SRAM cell}
    \label{fig:puf_basic_b}
  \end{minipage}
  \caption{Design illustration of Arbiter PUF and SRAM PUF.}
	\label{fig:puf_basics}
\end{figure}

\noindent$\bullet$ \textbf{PUF basics.} PUF is developed from the physical one-way function \cite{pappuPhysicalOneWayFunctions2002}. It is a hardware primitive that utilizes environmental variations in manufacturing and incorporates physical randomness into its structure. It can randomly map a given challenge $C$ to a response $R$ by such structure. For example, Arbiter PUF (APUF) design \cite{lee2004technique} (Figure \ref{fig:puf_basic_a}) compares the arrival delays of competing paths in an arbiter circuit to derive random responses. SRAM PUF design \cite{GuajardoSRAMPUF} leverages the start-up state of SRAM cells (Figure \ref{fig:puf_basic_b}) that exhibit random bit due to the MOSFETs difference.

Therefore, each PUF instance in the same design is unique and has a different CRP mapping, rendering physical duplication impossible. PUF can be divided into strong and weak PUF \cite{GuajardoSRAMPUF,RuhrmairPUFSP} in practice by the number of generated CRPs.



\noindent$\bullet$ \textbf{PUF in the real world.} Top MCUs and FPGA manufacturers have all developed their PUF designs \cite{nxplpcpuf, xilinxpuf}. Several PUF-specialized companies also emerged, such as Verayo, Intrinsic-ID \cite{intrinsicidpuf}, and ICTK \cite{ictkpuf}. PUFs have been quietly used in various real products, such as Intel SGX \cite{costan2016intel}, IoT \cite{pufiotrot}, medical wearables \cite{pufmedwearable} and SIM cards \cite{pufusim}. The academic research on PUF has drawn much attention these years \cite{chatterjeeBuildingPufBased2019, SGXFPGAXiaLX021,UnTrustZonePUF, MCUtokenPUF}. With ongoing standardization efforts \cite{PUFISOP1, PUFISOP2}, it is becoming a basic component in hardware ecosystems.


%








\noindent$\bullet$ \textbf{The benefits of PUF for CA} PUF can exploit true physical randomness to establish trust \textit{without keys}. Using it for key generation \cite{PUFSSL2018,PUFPKIIoT2022} does not provide stronger guarantees since it is possible to export the key in use. CAs still become compromised once that string gets exposed, regardless of how it was generated. We review the PUF's properties and discover that it can offer the following distinct advantages for CAs.





\ding{172} \textit{PUF is cryptographically sound.} Many studies \cite{PUFUCBrzuskaFSK11,ArmknechtPUFSP, RuhrmairPUFSP,DijkJPUFInterface} have confirmed the similarity between PUF and a random oracle or Pseudo-Random Function (PRF). The randomness and uniqueness of PUFs are highly desirable in cryptography since they are essentially the same as pseudo-randomness, one-wayness and other basic cryptographic properties. With proper abstraction, PUF can be a sound primitive \cite{xiaolinSPEAR} with physical obfuscations.

\ding{173} \textit{PUF can be a physical signing primitive for CA.} Based on \ding{172}, we further notice that PUF has a \textit{signing-like functionality} with physical authenticity to the output by its unique structure. It realizes the genuine binding between CAs and issued certificates. The traditional signature does not ensure \textit{who} actually holds the key. We do not intend to challenge these well-established algorithms, but think that PUF can be an alternative of generating CA endorsements. To some extent, it is both an intrinsic key and a signing primitive.








\ding{174} \textit{PUFs can be reliable in use.} The current (SRAM) PUF technology is quite mature for large-scale commercial use \cite{nxplpcpuf,ictkpuf,intelpuf,xilinxpuf}. Usually, a Error Correction Code (ECC) module is coupled with PUF circuits to stabilize the responses. As assumed in both industry and prior studies \cite{chaterjee3PAARivateUF2021,ArmknechtPUFSP,SGXFPGAXiaLX021,DijkJPUFInterface}, the ECC-backed PUF function for use can be 100\% stable under certain environmental changes. Moreover, PUF's aging time is long enough \cite{ROPUFAging} (\eg, years) to cover the certificate validity. The manufacturers will also comply with the ISO standards \cite{PUFISOP1,PUFISOP2} to further guarantee the performance and uniformity of the actual PUF hardware products.







PUF offers a distinct choice to prevent future exploitations. If we apply its ``keyless" signing functionality to the CA operations, the security risk of signing key exposure can be naturally eliminated since the explicit keys do not exist anymore.








\section{\acore\ Overview}\label{sec:overview}

In this paper, we propose \acore\ to fully reap the benefits of PUF for CAs. It does not aim to replace the public key cryptosystem but offers an alternative for more trusted certificate issuance in PKI.







\subsection{Threat Model}\label{sec:threat_model}

We assume a strong adversary $\adver$ who aims to expose the signing key of CAs, especially of the intermediate and online CAs, as they are more likely to be attacked. We categorize the attacks based on whether $\adver$ can reach to the CA server system.


\begin{itemize}
    \item \textbf{Remote attacks}: $\adver$ can conduct typical network attacks (\textbf{A1}\label{attack1}) in the Dolev-Yao model \cite{DolevYao}. For example, it can eavesdrop, intercept, modify and replay the certificates in the open channel. It can also abuse insecure configurations, implementation flaws or other vulnerabilities to export the key (\textbf{A2}\label{attack2}). Meanwhile, $\adver$ can conduct modeling attacks \cite{RuhrmairSSXMSDSBD13, KhalafallaEG20, WisiolTMSZ22} (\textbf{A3}\label{attack3}) against PUFs by collecting raw CRPs in the certificates. 
    \item \textbf{Onsite attacks}: $\adver$ may perform attacks against CA servers through a corrupted inside operator. It can install malicious components to directly copy out the signing key, or deduce the keys by non-invasive measures (\textbf{A4}\label{attack4}) such as (cache) side-channel analysis, software fault injection, etc. 
\end{itemize}

In the attacks above, $\adver$ aims to leak the key without attacking actual hardware. We do not consider the invasive physical attacks such as chip delayering or bus snooping since it is unrealistic to break the physical machines. Therefore, we trust the hardware stack of servers. Denial of Service (DoS) attacks are also excluded in this work. CAs invoke PUFs as they invoke other secure hardware. 


\acore\ focuses on \textit{preventing} the signing key exposure so it does not handle the misbehaviors of CAs \textit{after} they are totally controlled by $\adver$, which have been addressed by previous designs \cite{BasinARPKICKPSS14, MatsumotoRIKP17,ChuatKMPFPKI22, RHINEDuanFL0BP23}. Our design can well complement with them due to the orthogonal motivation and secure position.

\subsection{Design Overview}\label{sec:42_design_overview}



\begin{figure}[htbp]
    \centering
    \includegraphics[width=0.93\linewidth]{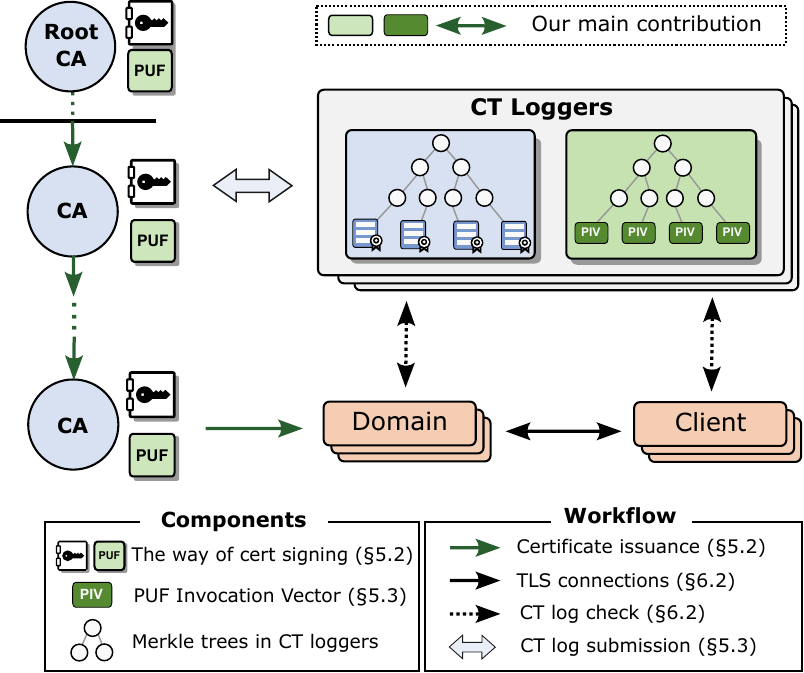}
    \caption{Architecture of \acore-enabled PKI.}\label{fig:acore_arch}
\end{figure}

\noindent$\bullet$ \textbf{System architecture.} The architecture of \acore\ is depicted in Figure \ref{fig:acore_arch}. It consists of four actors: CA, Domain, Client and Transparency Logger, which aligns with modern PKIs. In this work, we reuse the publicly trusted loggers of the existing CT infrastructure. We assume that the CA servers are attached with multiple PUF instances. Other actors maintain the same settings as usual. \acore\ does not introduce new entities or change the overall PKI workflow. The key difference lies in that CAs issue certificates (\cref{sec:5_certificate_issuance}) using PUF-based operations (\cref{sec:4_puf_certificate_ops}), and the clients will use PUF Invocation Vectors (PIVs) in the loggers (\cref{sec:4_puf_transparency}) to finish certificate verification (\cref{sec:5_certificate_validation}). 

\noindent$\bullet$ \textbf{Scope.} \acore\ does not intend to remove all keys in PKIs but only explores the physically trusted issuance for CAs through PUF-based operations. This enhanced PKI design can be combined with many other works \cite{BasinARPKICKPSS14,MatsumotoRIKP17, ChuatKMPFPKI22, Matsumoto20CAPSACSAC}. Most signing keys of other entities are still used in the rest procedures. \acore's focus is more on the intermediate and online CAs since the root CAs are quite secure on specialized, air-gapped machines. The hardware integration of PUF in commodity platforms are beyond the scope of this work.








\section{Building Blocks}\label{sec:acore_components}

We first introduce the essential building blocks of \acore\ that are the basis of the new PKI system.


\subsection{PUF Abstraction for CA Servers} \label{sec:puf_abstraction}


In previous works, PUFs are always used in resource-constrained platforms like endpoint devices. It is unclear about how they should be used in servers. Therefore, we need to clarify this in the paper.

\subsubsection{PUF Abstraction}\label{sec:subsec_puf_definition} Before fleshing out the formalized cryptographic definition, we describe the related notations as follows.

\noindent\textbf{Notations.} Let $\bin^{n}$ be the set of all $n$-bit strings, and $X \sample \bin^{n}$ means the random selection of a string $X$ from $\bin^{n}$. $\$(\cdot)$ is an $n$-to-$n$ random bits oracle that randomly maps input to the output string, \ie, $\$:\{0,1\}^{n}\rightarrow \{0,1\}^{n}$. The notation $|X|$ represents the length of $X$. For $X, Y \in \bin^{n}$, $X || Y$ denotes their concatenation, and $X \oplus Y$ is their bitwise XOR result. $H_{k}(\cdot)$ denotes a keyed cryptographic hash function, \ie, $H(k || \cdot)$. $\mathrm{Pr}[\cdot]$ is the probability of a given event. In the cryptographic context, we denote a theoretical distinguisher as $\Dd$, and $\Dd^{(\cdot)} \Rightarrow 1$ means that $\Dd$ successfully distinguishes a cryptographic oracle $(\cdot)$ by querying its output.

\begin{definition}($\epsilon$-secure PUF)\label{def:puf_definition}
    Let $\Pp$ be a PUF family where the response $R=\puf{}(C)$ for $ \forall \puf{} \in \Pp$ and $C, R \in \{0, 1\}^{n}$. $\Pp$ is called $\epsilon$-secure if for any distinguisher $\Dd$, we have,
    \begin{itemize}[itemsep=0pt]
		\item (Randomness) $\forall \puf{} \in \Pp, \forall C \in \{0,1\}^{n}$, 
		\begin{equation}
			\label{eq:ideal-PUF-pseudorandomness}
			\left|\mathrm{Pr}\left[\Dd^{\puf{}(C)}\!\Rightarrow 1\right]\! - \!\mathrm{Pr}\left[\Dd^{\$(C)}\!\Rightarrow 1 \right]\right| \le \frac{1}{2^{n}}\cdot \epsilon(n),
		\end{equation}
		where $\epsilon(n)$ is a negligible value related to $n$. Formula \ref{eq:ideal-PUF-pseudorandomness} means that $\Dd$ can only distinguish a PUF oracle from $\$(\cdot)$ with negligible probability, which suggests that the probability distribution of PUF responses are almost identical to the true random bits.
		\item (Inter-Uniqueness) $\forall \puf{}_{1}, \puf_{2} \in \Pp, \forall C \in \{0,1\}^{n}$, let $p(C)=\puf{}_{1}(C)\allowbreak \oplus \puf_{2}(C)$, 
		\begin{equation}
			\label{eq:ideal-PUF-uniqueness}
				\left|\mathrm{Pr}\left[\Dd^{p(C)}\Rightarrow 1\right]-\mathrm{Pr}\left[\Dd^{\$(C)}\Rightarrow 1\right]\right| \le \frac{1}{2^{n}}\cdot \epsilon(n).
		\end{equation}
        Formula \ref{eq:ideal-PUF-uniqueness} indicates that the inter-Hamming distance of different PUF instances is large enough so that each PUF instance and the responses can be seen as statistically independent.
        \item (Intra-Uniqueness) $\forall \puf{} \in \Pp$, $c \sample \{0,1\}^{n}$ and $\forall C_{1},C_{2} \in \bin^{n}, C_{1}\allowbreak \ne C_{2}$,
		\begin{equation}
			\label{eq:ideal-PUF-axu}
			\mathrm{Pr}\left[\puf{}(C_{1})\oplus \puf{}(C_{2})=c\right] \le \frac{1}{2^{n}}\cdot\epsilon(n).
		\end{equation}
		This property demonstrates that the responses for one PUF instance are also uniformly distributed. 
        \item (Unclonability and Tamper Resistance) Every $\puf{}\in\Pp$ has a unique physical structure. Any invasive measures to probe the PUF instance would alter its structure and turn it ineffective or into a different instance.
    \end{itemize}
\end{definition}

As discussed in \cref{sec:3_puf_advantage}, ECC modules are default coupled with PUFs to ensure the reliability in use. So we omit them in the definition to focus on these properties. We regard PUF as a static and unique entropy source that remains stable under challenges, which aligns with the previous studies \cite{ArmknechtPUFSP, delvaux2017security,chatterjeeBuildingPufBased2019,xiaolinSPEAR} and the standards \cite{PUFISOP1, PUFISOP2}. Any PUF design that matches this definition can be used here.



\subsubsection{Usage of PUF in CAs}\label{sec:subsec_puf_usage_ca}

For resource-rich platforms like CA servers, it is reasonable to install multiple PUF instances due to their low cost and energy efficiency \cite{PUFarbiterperf2004,pufusim}. Therefore, unlike endpoint devices, for a CA with name $N_{CA}$, we make it to be equipped with a \textit{group} of $\epsilon$-secure PUFs $P=\{\puf_{i}, \dots, \puf_{m}\}$ and $m$ can be hundreds or more. This approach allows CAs to decouple the specific PUF implementations and the PKI functions that rely on them, thereby enabling the integration of different PUFs. Furthermore, it has additional benefits in practice that have never been explored before.

\begin{itemize}
    \item \textbf{PUF parallelization.} CAs can invoke multiple PUFs in parallel to increase the throughput of PUF-related operations.
    \item \textbf{Randomized PUF rotation.} CAs can randomly select one PUF instance from the group to use. This can avoid the fixed calling pattern to mitigate PUF modeling attacks \cite{WisiolTMSZ22,RuhrmairSSXMSDSBD13}.
\end{itemize}




Such abstraction provides a new usage model of PUFs in server-like environments. CAs then can treat PUF as a keyless signing primitive to sign certificates without using an explicit key.




\subsection{PUF-based X.509 Certificates}\label{sec:4_puf_certificate_ops}

\begin{figure}[htbp]
    \centering
    \includegraphics[width=0.91\columnwidth]{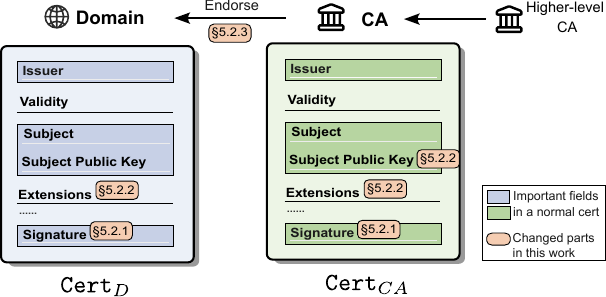}
    \caption{Changed fields of PUF-based certificates.}
    \label{fig:puf_certs}
\end{figure}

Based on the abstraction above, CA can use $\puf{} \in P$ to create \textit{physically} trusted binding of a domain to its public key $pk_{D}$ in the issued certificate $\texttt{Cert}_{D}$. However, this would change the core logic of traditional PKI systems. To address this challenge (\textbf{Ch1}\label{challenge1}), we design the PUF-based certificates with minimal modifications to the X.509v3 format. No other fields are changed except for the annotated ones in Figure \ref{fig:puf_certs}. In the rest of this paper, without loss of generality, we consider a multi-level chain of trust with one root CA $N_{CA_{0}}$, $M$ intermediate CAs $N_{CA_{M}}$ and a domain $N_{D}$.





\subsubsection{Physically Trusted Signature}\label{sec:41_cert_sig_field}

\begin{figure}[htbp]
    \centering
    \includegraphics[width=0.96\columnwidth]{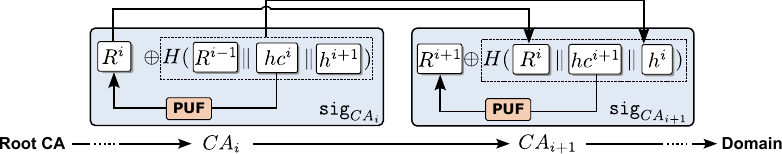}
    \caption{Signature structure of PUF-based certificates.}\label{fig:cert_sig_chain}
\end{figure}

The signature field $\texttt{sig}$ of a certificate is usually calculated as $\texttt{sig}=\textsc{Sign}(\texttt{crt})$ where $\textsc{Sign}$ is a standard signing algorithm, and $\texttt{crt}$ denotes to-be-signed certificate entries, including the subject name, public key, issuer name, etc. In this work, $\texttt{sig}$ has a chained structure, which is composed of a PUF response and hash chain pointers as shown in Figure \ref{fig:cert_sig_chain}. 

For a domain $N_{D}$, the signature in $\texttt{Cert}_{D}$ is $\texttt{sig}_{D}=R_{D} \oplus H(R_{CA} ||$\\$ hc_{D} || h_{CA})$ where $hc_{D} = H(\texttt{crt}_{D}||\texttt{ts}||N_{CA})$ and $R_{D}=\puf{}\ (hc_{D})$. \texttt{ts} is the timestamp of invoking \puf{}, and $N_{CA}$ denotes the name of the issuing CA. This structure is also applied to the whole chain of trust of CAs: if $\texttt{sig}_{D}$ is assumed to be $\texttt{sig}^{*}$ in Figure \ref{fig:cert_sig_chain}, then $\texttt{sig}^{+1}$ is the signature in $N_{CA}$'s own certificate and so on.


In our design, PUF directly offers physical authenticity and integrity to the certificate entries, which is different from key-based approaches like PUFSSL \cite{PUFSSL2018}. The chain of $\texttt{sig}$ serves as physically trusted proofs of genuinely conducting the issuance by the CAs. 


\subsubsection{Public Inclusion Proof}\label{sec:41_inclusion_proof}

With PUFs being used for signing, it is necessary to provide a public proof in certificates to show that the used PUF instance is authentic, \ie, belongs to that CA. To this end, we design the PUF inclusion proof mechanism, which would change the public key and extension fields.

\noindent$\bullet$ \textbf{Public key field.} For domains, the public keys $pk_{D}$ remains unchanged. For CAs, $\texttt{pk}$ field in $\texttt{Cert}_{CA}$ is set to $\pi_{CA}=\oplus_{i=1}^{m} \puf_{i}(C_{CA})$ where $\puf_{i}$ are the installed PUF instances, and $C_{CA}$ is a unique identity challenge for each CA. $\pi_{CA}$ combines the identity response $\puf_{i}(C_{CA})$ from each PUF instance to form the unique public proof associated to a private credential, \ie, PUF, of that CA. This is why we use it to set \texttt{pk}. However, unlike a public key, $\pi_{CA}$ is not used for the actual verification of a PUF-based signature, but helps confirm the authenticity and ownership of the used PUF instance. 

$\pi_{CA}$ is calculated by CA itself through XOR-based accumulation. CA can update it by simply XORing the corresponding response to $\pi_{CA}$ when there are new PUF instances or removed ones, just like updating the public key when the private key is changed.

\noindent$\bullet$ \textbf{Extension field.} There are reserved X.509v3 extension fields $\texttt{ext}$ for certificate customization. In this work, we set $\texttt{ext}$ to the identity response $RP_{k}=\puf{}_{k}(C_{CA})$ where $\puf_{k}$ is the PUF instance used by the issuer to generate \texttt{sig}. Its complementary part $\widetilde{RP_{k}} = \pi_{CA} \oplus RP_{k}$ is maintained by the loggers, as detailed in \cref{sec:4_puf_transparency}. Note that $RP_{k}$ and $\pi_{CA}$ are not in the same certificate but located separately in the trust chain. For example, if $RP_{k}$ is set in the domain certificate $\texttt{Cert}_{D}$, then the corresponding $\pi_{CA}$ is actually the \textit{issuer}'s \texttt{pk}.

This design ensures that \texttt{pk} in CA's certificate is unchanged while the certificates it issues can contain different $RP_{k}$ to identify the used PUF. This allows verification of whether a PUF instance is included in the CA's PUF group. When a verifier (\eg, a client) resolves the certificate chain and retrieves $\widetilde{RP_{k}}$, it can check if $\widetilde{RP_{k}} \oplus RP_{k}$ equals to that $\pi_{CA}$ before verifying the signature (\cref{sec:5_certificate_validation}).




\subsubsection{PUF-based Certificate Chain}\label{sec:certificate_chain}

\begin{figure}[htbp]
    \centering
    \includegraphics[width=0.89\columnwidth]{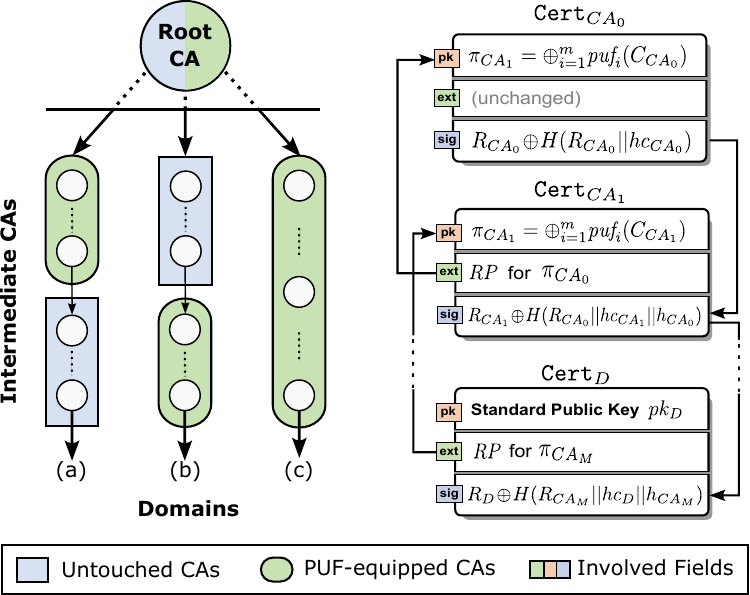}
    \caption{PUF-based CA chain and certificate chain.}
    \label{fig:puf_certs_chain}
\end{figure}

Now we put these designs together in the multi-level certificate chain shown in Figure \ref{fig:puf_certs_chain}. The PUF-based trust chain can coexist with the original certificates. The root CA $N_{CA_{0}}$ is modified only because a starting point is needed in the PUF-based trust chain. Since it operates in highly secure, completely isolated air-gapped environments, removing its signing key is not necessary. Consequently, clients need to install a PUF-based root certificate $\texttt{Cert}_{CA_{0}}$ in company with the original one. 

In $\texttt{Cert}_{CA_{0}}$, $\texttt{sig}$ is generated by $N_{CA_{0}}$'s own PUF, which is similar to the self-``signing'' operation. Specifically, we have $\texttt{sig} = R_{CA_{0}} \oplus h_{CA_{0}}$ where $h_{CA_{0}}=H(R_{CA_{0}} || hc_{CA_{0}})$ and $R_{CA_{0}}=\puf{}\ (hc_{CA_{0}})$. This would initiate the construction of our signature chain, as depicted in Figure \ref{fig:cert_sig_chain}. As for the $\texttt{pk}$ and $\texttt{ext}$ fields, they are set to the corresponding $\pi_{CA}$ (upper level) and $RP_{k}$ (lower level), as illustrated in the right part of Figure \ref{fig:puf_certs_chain}. Note that such transformation is not mandatory for every CA. CAs can freely choose whether to adopt this PUF-based chain to benefit from keyless security.


\noindent$\bullet$ \textbf{Interoperability with traditional certificates.} The PUF-based certificate chain can be interoperable with the original CA. As shown in the left part of Figure \ref{fig:puf_certs_chain}, we consider three possible cases:

\begin{itemize}
    \item \textbf{(a)} When PUF-equipped CAs are \textit{issuers}, the issued certificates will contain PUF-based signatures (\cref{sec:41_cert_sig_field}). The $\texttt{pk}$ field can remain unchanged to issue standard certificates.
    \item \textbf{(b)} When PUF-equipped CAs are \textit{issuees}, their public keys now are the $\pi_{CA}$ (\cref{sec:41_inclusion_proof}), but the $\texttt{sig}$ field can be the traditional signature, depending on the upper-level CA's choice.
    \item \textbf{(c)} In this case, the entire certificate chain is depicted on the right part where the three fields are transformed accordingly.
\end{itemize}

Overall, the new certificates modify at most three fields while adhering to other X.509 specifications (\hyperref[challenge1]{\textbf{Ch1}}). It maximumly preserves compatibility although inevitably introducing certain changes.










\subsection{PUF Transparency}\label{sec:4_puf_transparency}

Now we can enable physically trusted certificate issuance for CAs, but how to verify the certificates remains a challenge because user agents cannot interact with CAs to query PUF again (\textbf{Ch2}\label{challenge2}). In a naive approach, they must somehow pre-store enough CRPs for verification, but this is impractical because clients cannot foresee which domain to authenticate. And the extra CRP storage can cause significant overhead. Besides, there is currently no way to monitor the PUF usage in CAs to mitigate the potential abuse (\textbf{Ch3}\label{challenge3}).



To address this challenge, we propose PUF transparency mechanism. It can be realized on a specialized trusted server that is common in previous designs \cite{Matsumoto20CAPSACSAC, chatterjeeBuildingPufBased2019,chaterjee3PAARivateUF2021,zhengPUFbasedMutualAuthentication2022}. This may require new trust assumptions and workflow modifications although it eases the design. In this work, we opt to leverage the existing CT infrastructure for seamless integration. Here we follow the Google CT policy \cite{googleCTpolicy} which stipulates that loggers $\lgr$s only accept submissions from the pre-defined CAs.


 

\subsubsection{PUF Invocation Vector}

The core of this mechanism is the PUF Invocation Vector (PIV), a new data structure to record the calling behaviors of PUFs. Specifically, we define a PIV as $I=\langle Z, T\rangle$. The description tuple is $Z=\{N_{CA}, N_{P}, \texttt{ts}, \widetilde{RP_{k}}\}$. It means that the CA $N_{CA}$ invoked PUF that is produced by the manufacturer $N_{P}$ at the time $\texttt{ts}$. $\widetilde{RP_{k}}$ is the complementary proof (\cref{sec:41_inclusion_proof}). The authenticated tag $T=H_{R_{CA}}(Z||\texttt{Cert}||T_{h})$ where $R_{CA}$ is the response used in \texttt{sig} and $T_{h}$ is the tag of the higher-level CA's PIV. 

\begin{figure}[htbp]
    \centering
    \includegraphics[width=0.93\columnwidth]{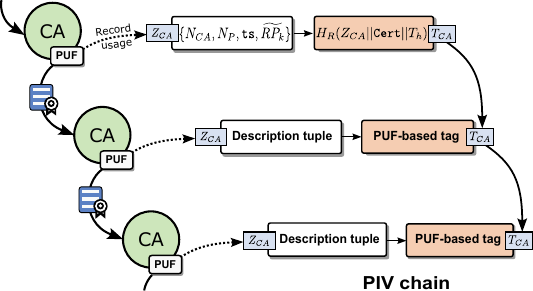}
    \caption{Chain Structure of PUF Invocation Vectors.}\label{fig:hash_chain}
\end{figure}

Each time CA invokes a PUF instance for issuance, it will generate a corresponding PIV. These PIVs would form a similar chain related to the PUF-based certificate chain as shown in Figure \ref{fig:hash_chain}.


In the chain of PIVs, the nested hash tag offers a HMAC-like functionality to ensure the integrity. This structure binds certificates and PUF invocation behaviors through the unique PUF response.

\subsubsection{Integration to CT}\label{sec:4_puf_transparency_CT_integration}

We now explain how to integrate PIVs into CT infrastructure to enable PUF transparency. 

\noindent$\bullet$ \textbf{Submitting PIVs}. CAs can upload PIVs along with certificates to $\lgr$s. The loggers can employ another Merkle Tree to log PIVs in the same manner as certificates. The introduction of PIVs does not change the append-only functionality of CT. The monitors \cite{rfc9162} in the CT ecosystem ensure that these logs are not tampered with and will report any suspicious abuse behavior.

\noindent$\bullet$ \textbf{Delivering PIVs}. To avoid extra interactions of delivering PIVs to clients for verification, we use the same approach as SCTs (Signed Certificate Timestamps) in CT. Specifically, we leverage OCSP stapling mechanism by which clients can receive SCTs and PIVs together when they check the revocation status of the received certificates. Also, it has been supported by real-world projects like OpenSSL and Nginx, enabling PIVs to enjoy this established infrastructure. Note that clients need to retrieve PIVs through a signed OCSP response immediately upon receiving the certificates. 


This PUF transparency mechanism enables the public audits of PUF invocation and facilitates efficient verification (detailed in \cref{sec:5_certificate_validation}), thereby addressing \hyperref[challenge2]{Ch2} and \hyperref[challenge3]{Ch3}. To the best of our knowledge, it has not been discussed by previous PUF-based works \cite{chatterjeeBuildingPufBased2019,zhengPUFbasedMutualAuthentication2022,zhang2024teamwork}.

\section{PKI Workflow in\ \acore}\label{sec:acore_functions}

\begin{figure}[htbp]
    \centering
    \includegraphics[width=\columnwidth]{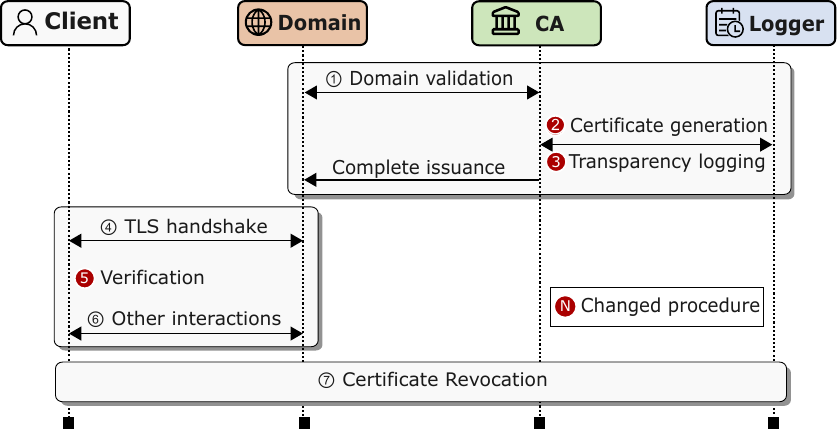}
    \caption{Complete PKI Workflow with \acore.}\label{fig:acore_pki_protocol}
\end{figure}

Now with the building blocks above, we give the complete workflow in an \acore-enabled PKI system, as shown in Figure \ref{fig:acore_pki_protocol}.




\subsection{Certificate Issuance}\label{sec:5_certificate_issuance}

$\bullet$ \textbf{Domain validation} (\ding{172}). This step is untouched and might not be necessary for all PKIs. The CA can use standard HTTP- or DNS-based protocols to check the domain's authority before the issuance.

\noindent$\bullet$ \textbf{Domain certificate generation} (\ding{173}).  We assume that CAs have prepared their own certificates, \ie, from $\texttt{Cert}_{CA_{0}}$ to $\texttt{Cert}_{CA_{M}}$. Next, as described in \cref{sec:puf_abstraction} and \cref{sec:4_puf_certificate_ops}, $N_{CA_{M}}$ randomly invokes an PUF instance to generate $\texttt{sig}_{D}$. It assembles the final certificate $\texttt{Cert}_{D}$ to form the complete chain in Figure \ref{fig:puf_certs_chain}. The domain receives $\texttt{Cert}_{D}$ as a trusted endorsement of its public key from $N_{CA_{M}}$.

\noindent$\bullet$ \textbf{Transparency logging} (\ding{174}). $N_{CA_{M}}$ generates the corresponding PIV $I_{D}$ (\cref{sec:4_puf_transparency}) to record the above PUF calling behavior. Then it uploads $\texttt{Cert}_{D}$ and $I_{D}$ to one of the loggers $\lgr$s for transparency logging. These distributed loggers under continuous monitoring by other CT actors \cite{rfc9162} will synchronize with each other to achieve the tamper-evident trusted logging.

\subsection{Certificate Verification}\label{sec:5_certificate_validation}

We specify the certificate verification function of \acore\ in Algorithm \ref{alg:certificate_validation}. It follows the common paradigm of most PUF-based designs \cite{chatterjeeBuildingPufBased2019,zhengPUFbasedMutualAuthentication2022,SGXFPGAXiaLX021,PUFPKIIoT2022}, which requires the implicit comparison between the ``golden value'' and the actual PUF response.  


\noindent$\bullet$ \textbf{Retrieving the certificate and PIV} (\ding{175}). As described in \cref{sec:4_puf_transparency_CT_integration}, when a client receives the certificate of a domain $N_{D}$ through TLS handshake, it will request the SCTs and PIVs to be delivered together through OCSP stapling \cite{rfc9162}. The client can resolve the whole certificate chain and PIV chain (line \ref{alg:line1} and \ref{alg:line2}). Here we assume a linear certificate chain for simplicity, although the organization of the real certificates may be complex in practice. 

\noindent$\bullet$ \textbf{Verifying the certificate} (\ding{176}). Next, the client will start the verification from the root certificate to the domain certificate. It parses out the certificate entries \texttt{crt} (line \ref{alg:line5}) and recalculates $hc$ (line \ref{alg:line6}). Then it can recover the correct "signature" responses $R_{CA}$ and $R_{D}$ (line \ref{alg:line15}) only if the entries and signatures are not tampered. The verification of our chaining signatures (\cref{sec:41_cert_sig_field}) is symmetric to their constructions due to the use of hash and XOR operations. The tags in PIVs are also recalculated and checked via the signature responses (line \ref{alg:line16}).




\begin{algorithm}[htbp]
    \fontsize{9}{9}
    \setstretch{1.05}
    \caption{Certificate Verification Algorithm}\label{alg:certificate_validation}
    \SetNoFillComment
    \Input{Certificate chain \textsf{Certs}, PIV chain $\mathcal{P}IVs$}
    \Output{Validation result (\textsf{valid} or \textsf{invalid})}
    $\textsf{Certs} = \{\texttt{Cert}_{CA_{0}}, \texttt{Cert}_{CA_{1}}, \cdots, \texttt{Cert}_{CA_{M}}, \texttt{Cert}_{D}\}$\; \label{alg:line1}
    $\mathcal{P}IVs=\{I_{CA_{0}},I_{CA_{1}},\cdots,I_{CA_{M}},I_{D}\}$\label{alg:line2}\;
    \textbf{Init} $h_{last}, R_{last}, T_{last}, \pi_{last}=$ \textbf{Null}\;
    \For(\tcp*[f]{Main verification loop}){\textsf{cert}, $I$ \textbf{in} $\textsf{Certs}, \mathcal{P}IVs$}{
        $Z, T = I.\texttt{parse()},\ \texttt{crt} = \textsf{cert}.\texttt{parse()}$\;\label{alg:line5}
        $hc = H(\texttt{crt} || Z.\texttt{ts} || Z.N_{CA})$\label{alg:line6}\;
        \eIf{\textsf{cert}.\texttt{name} \textbf{is} ``RootCA''}{
            $h=hc$\;
        }{
            $h=H(R_{last} || hc || h_{last})$\;
            \If{$RP$ \textbf{in} $\textsf{cert}.\texttt{ext}$ \textbf{and} $RP\oplus Z.\widetilde{RP_{k}}\neq\pi_{last}$}{
                \Return{\textsf{invalid}}; \tcp{PUF inclusion proof} 
                
            }\label{alg:line12}
        }
        $R=h\oplus \textsf{cert}.\texttt{sig}$; \tcp{Recover the response}\label{alg:line15}
        $T'= H_{R}(Z || \textsf{cert} || T_{last})$; \tcp{The chained PIV tag}\label{alg:line16}
        \If{$T' \neq T$}{
            \Return{\textsf{invalid}}; \tcp{Check both chains}\label{alg:line18}
        }
        \tcc{Update the variables for the next level}
        $h_{last} = h, R_{last} = R, T_{last} = T'$\; 
        $\pi_{last}=\textsf{cert}.\texttt{pk}$\;
    }
    \Return{\textsf{valid}}\;
\end{algorithm}

The client performs the same steps to verify the chain of certificates and PIVs all the way down from the root CA to the domain. The whole verification process succeeds only if the two chains are both authentic in the end (line \ref{alg:line18}). In the signature verification, the public proofs of PUF (\ref{sec:41_inclusion_proof}) are primarily checked by the $\texttt{pk}$ and $\texttt{ext}$ field (line \ref{alg:line12}). It is not directly involved in verifying the signature itself like a standard public key. Once they are verified, the client can proceed with the rest procedure (\ding{177}) if all the checks pass. The regular checks of the expiration date, policy constraints and others are omitted in Algorithm \ref{alg:certificate_validation}.

As presented in Algorithm \cref{alg:certificate_validation}, PIVs from $\lgr$s now are directly used in the actual signature verification. Their authenticity and integrity are mainly ensured by the CT ecosystem in \cref{sec:4_puf_certificate_ops}, which is indeed different from traditional public key algorithms. However, we must emphasize that this does not simply shift the trust from CAs to CT. The root of trust in certificates is still the CA's (PUF-based) endorsement. Only part of the trust in the verification process are now spread to the whole CT ecosystem. Its threat model remains unchanged, and the PIVs does not require higher guarantees than the logged certificates or SCTs.

\subsection{Certificate Revocation}\label{sec:5_certificate_revoke}

In \acore, the introduction of PUF does not affect the lifetime and expiration date fields. So it is compatible with common revocation techniques such as OCSP and CRL (\ding{178}). However, when a PUF instance is failed due to hardware glitches or natural aging, the certificates recently issued by it should be revoked as well. Since CAs is equipped with multiple PUF instances, they can use other functional instances for issuance and replace the revoked ones later.

\subsection{Deployment Considerations}\label{sec:deployment_considerations} \acore\ is designed to be compatible with global PKIs, \eg, Web PKI, code-signing PKI and resource PKI. However, considering the inconveniency of hardware modification, we do not expect it to be deployed instantly like previous works \cite{Matsumoto20CAPSACSAC,ChuatKMPFPKI22}. It can be used first on a small scale, such as in private PKIs. It does not require a ``flag'' day to be effective and CAs will enjoy the security gains immediately after the deployment.

\acore\ does not restrict how the PUF hardware should be coupled with the CA servers. They can be attached like a standalone HSM card, or be integrated into central processors as a security extension like TEE. As for the software part, it can co-exist with traditional PKI functions to be deployed incrementally on the fly without the global replacement of client software.

\section{Security Analysis}\label{sec:security_analysis}

\begin{figure}[htbp]
    \centering
    \includegraphics[width=0.89\columnwidth]{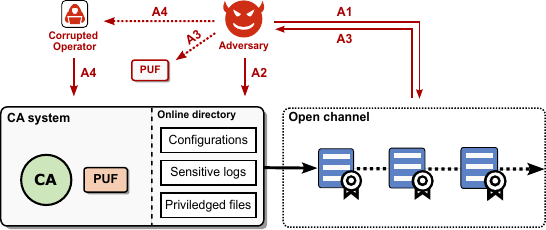}
    \caption{Attack view illustration in \acore.}
    \label{fig:attack_views}
\end{figure}

In this section, we analyze the security of our design on the threat model (\cref{sec:threat_model}). We present the possible attack views in Figure \ref{fig:attack_views}.

The overall security goal of \acore\ is to eliminate the risk of exposing a CA's signing key against the strong adversary $\adver$. Many common attack vectors are automatically eliminated thanks to the PUF-based keyless issuance. $\adver$ will fail with \hyperref[attack2]{\textbf{A2}} because the ultimate target, \ie, CA's signing key, does not exist in code, storage, configs or any other places. The human errors cannot cause the key leakage anymore, either. Consequently, $\adver$ will also fail even with the help of an corrupted insider (\hyperref[attack4]{\textbf{A4}}).

Therefore, we will focus on the other two adversarial objectives in the following: certificate forgery (\hyperref[attack1]{\textbf{A1}}) and PUF modeling (\hyperref[attack3]{\textbf{A3}}).



\noindent$\bullet$ \textbf{Channel attacks (\hyperref[attack1]{\textbf{A1}})}: If $\adver$ aims to forge fields like \texttt{sig} and $\pi_{CA}$ of the issued certificates or PIVs, it must be able to predict PUF responses of arbitrary challenges, and break the hash function. However, under the $\epsilon$-secure PUF definition, $\adver$ cannot achieve this goal which is rigorously proved with the full cryptographic proof in Appendix \ref{sec:full_proof}. The security of our certificate construction is bounded by the security of underlying PUF.

$\adver$ can also resolve a certificate chain to recover the chain of PUF responses, \ie, $R_{CA}$ or $R_{D}$. It could strip off these responses and replace them with self-generated ones to create fake certificates. However, to pass the verification, $\adver$ would need to forge the corresponding SCTs, PIVs, and their inclusion proofs in CT, which essentially compromises the entire CT infrastructure. Those distributed, frequently synchronized loggers and other active monitors ensure that suspicious modifications can be detected. Besides, we have prevented the unauthorized log submission from the beginning \cite{googleCTpolicy}. Therefore, if $\adver$ wishes to forge a valid certificate, it have to break the PUF security remotely or compromise the widely served CT ecosystems in PKI.







\noindent$\bullet$ \textbf{PUF modeling attack (\hyperref[attack3]{\textbf{A3}})}: Although PUF greatly reduces the attack surface, the introduction of it can bring a new threat: PUF modeling attacks \cite{KhalafallaEG20, WisiolTMSZ22}. $\adver$ may exploit some automatic CA interfaces to acquire numerous certificates, thereby recovering enough CRPs to train a PUF model by machine learning. However, this attack is less of concern for our design. First, we need to point out that the cost of such attack is quite high in PKI scenarios because $\adver$ cannot directly probe the PUF hardware \cite{RuhrmairSSXMSDSBD13}. And it may require plenty of CRPs, \eg, $>\!100$K or even millions \cite{XuTransferResilientPUF} to train a 128/256-bit strong PUF model for one specific instance.

Second, given this condition, \acore\ is built on the proposed PUF abstraction model (\cref{sec:puf_abstraction}) where the PUF instances are randomly selected for use. This mechanism fundamentally breaks the prerequisite for modeling attacks. The CRPs that $\adver$ collects are not from the same instance. When they are generated from different PUF architectures, there is currently no method to train a unified model for all PUF designs. Even if there is a CRP-limited PUF instance used by the CA, $\adver$ cannot differentiate its CRPs as the PUF calling pattern has been hidden. Besides, the design of secure PUFs has been a prosperous research area and there are many PUFs with enhanced modeling resilience \cite{XuTransferResilientPUF,NassarANVPUF, LinZPZPUF} that can be used in \acore\ due to our PUF-agnostic design.

\section{Implementation}\label{sec:implementation}

We implement two open-sourced prototypes \cite{acoregithub} for \acore: \ding{172} a software functionality prototype that provides main PKI functions for performance evaluation, and \ding{173} a hardware prototype, a customized CPU with the built-in PUF module to demonstrate the hardware overhead and deployment feasibility. 

\begin{table}[htbp]
    \centering
    \caption{Components of our functionality prototype.} \label{tab:acore_impl_component}
    \fontsize{8pt}{8pt}\selectfont
    \begin{tabular}{clcc}
        \toprule
        \textbf{Used by} & \textbf{Code Base} & \textbf{Language} & \textbf{LoC} \\
        \midrule
        CA & \texttt{Let's Encrypt Pebble}, \texttt{gRPC} & Golang & 1.9K \\
        Domain & \texttt{EFF Certbot}, \texttt{gRPC} & Python & 0.4K \\
        Logger & \texttt{Google Trillian}, \texttt{gRPC} & Golang & 0.3K \\
        Client & \texttt{Mozilla WebExtension}, \texttt{pbf} & Javascript & 0.6K \\
        \bottomrule
    \end{tabular}
\end{table}

\noindent$\bullet$ \textbf{Functionality prototype.} Table \ref{tab:acore_impl_component} summarizes the components of the software prototype. We have developed the PKI functions of \acore\ in Pebble \cite{pebble}, a simplified version of Let's Encrypt CA Boulder for test use. We replaced \texttt{x509.CreateCertificate} function with the PUF-based function \texttt{CreatePUFCertificate} in the new \texttt{pcert} Golang package. We simulated a hash-based PUF module in it to complete the workflow, which can be replaced by actual PUF hardware in the future. The general issuance logic in Pebble remains unchanged but we added PUF-related operations including PIV creation, PUF inclusion proof generation, etc.

For Certbot, we implement a new function \texttt{acore\_verify} that has customized parsing and verification functions. It can be used by domains to interact with PUF-enabled CA in the ACME protocol. Additionally, we created a browser extension with Mozilla Web Extension API. Certain X.509 parsing functions were hand-coded in JavaScript to enable the customized modification to the verification logic. The PUF transparency mechanism is built upon Google Trillian \cite{trillian}, an industry-validated project used as the back-end of CT. We did not modify Trillian itself but developed new functions like PIV uploading and checking via its API. 

We employed \texttt{protobuf} to achieve consistent data serialization. These four components can then assemble and parse the certificates and PIVs in a unified format among each other.





\noindent$\bullet$ \textbf{Hardware prototype.} To demonstrate the feasibility of enabling PUF in CA servers. We embedded an IPUF \cite{InterposePUF19CHES} instance, a modeling-resistant strong PUF into CVA6 \cite{RISCVCVA6}, a 6-stage RISC-V CPU. We developed it using Vivado 2018.2 with $>$43K LoC on FPGA. This CPU can provide a set of native PUF RISC-V instruction extensions, \texttt{XPUF}, to control and interact with the hardware PUF module for the hosted applications. With \texttt{XPUF}, it becomes feasible to aggregate multiple PUF resources, thereby laying the foundation for heterogeneous platforms. More details are provided in Appendix \ref{sec:riscv_platform}.



\begin{figure*}[htbp]
    \centering
    \includegraphics[width=0.95\textwidth]{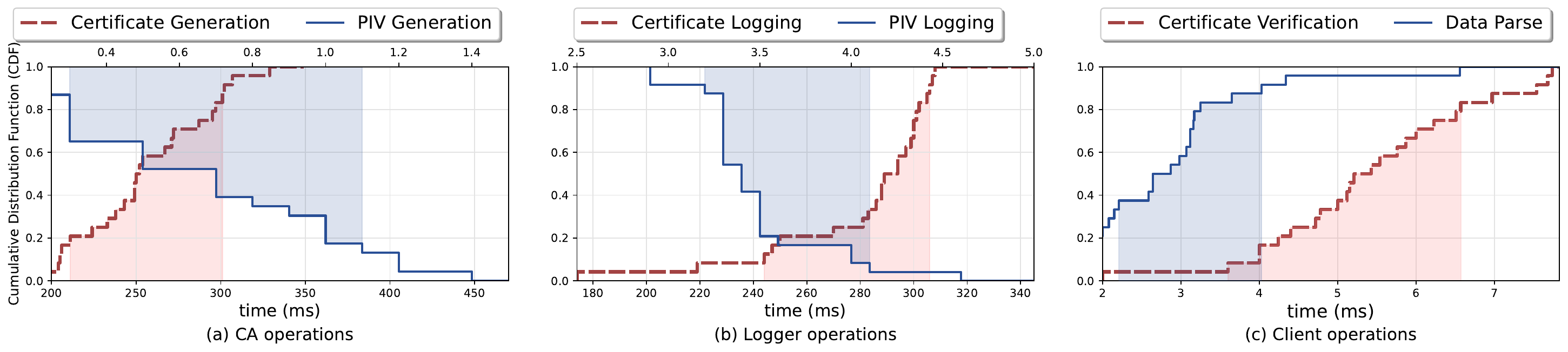}
    \caption{Performance breakdown of \acore\ functionality prototype.}
    \label{fig:q1-ops}
\end{figure*}

\section{Evaluation}\label{sec:evaluation}

In a real-world complex PKI workflow, there are numerous evaluation factors that can be considered. In this section, we prioritize answering the following questions. 


\begin{itemize}[itemsep=0pt, parsep=0ex]
    \item[\textbf{Q1}]: How does PUF compare to common signing primitives in certificate issuance? How is \acore\ different from other designs? 
    \item[\textbf{Q2}]: What is the detailed performance of the \acore\ prototype? 
    \item[\textbf{Q3}]: How much overhead does \acore\ introduce to the original PKI system? 
    \item[\textbf{Q4}]: How much does PUF and \acore\ cost for additional hardware deployment?
\end{itemize}

\textbf{Testbed.} For our functionality prototype, the code of CA, transparency loggers and domain are deployed on a Dell EMC R750 Server running Debian GNU 11 with 512 GB RAM and Intel Xeon Gold 6330 CPU@3.1 GHz. It is SGX-enabled and SDK version is 2.19. The client browser extension is integrated in Firefox v124.0.2 on a laptop with 16 GB RAM and Intel Core i5-8279U CPU@2.4 GHz. For the RISC-V hardware prototype, it is evaluated on the Digilent Genesys-2 Kintex-7 FPGA development board, which is based on 28nm Xilinx FPGA technology with maximum clock frequency of 1286.0 MHz. These components can communicate with each other through a local network with $\sim$1 Gbps bandwidth. 

In the following, we will run each test for 30 times and report the average results. Also, since the evaluation of PUF parallelization is beyond our focus, we only use one PUF instance in both prototypes.




  




\subsection{Q1: Comparison}\label{sec:eval_puf_primitive}

\textbf{Primitive comparison.} First, we investigate the performance of the core part of our design, \ie, the PUF ``signing'' operation. To cover a wide range of various PUFs, we present six representative designs in Table \ref{tab:puf_evaluation}. We can see that most of latencies typically fall in the millisecond range, which is quite performant considering its low manufacturing cost and moderate oscillating frequency. 


\begin{table}[htbp]
    \centering
    \begin{threeparttable}
    \caption{Performance of representative PUF designs.} \label{tab:puf_evaluation}
        \fontsize{7pt}{7pt}\selectfont
        \begin{tabular}{ccccccc}
            \toprule
            \textbf{Designs} & \cite{AnandaPUF} & \cite{PUFCOTE22} & \cite{GunluROPUF} & \cite{InterposePUF19CHES} & \cite{FEFETPUFGuo21} & PUF$_{6}^{\dagger}$ \\
            \midrule
            \textbf{Type} & Weak & Strong & Weak & Strong & Weak & Strong \\
            \textbf{Technique} & Latch & Arbiter & RO & Arbiter & FeFETs & - \\
            \textbf{Energy} & 0.7 pJ/b & - & - & - & 1.75 fJ/b & - \\
            \textbf{Latency} & 1 ms & 0.8 ms & 1.7 ms & 5.6 ms & 0.8 ms & 3 ms \\
            \textbf{Length} & 256 & 128 & 1275 & 128 & 256 & 256 \\
            \textbf{Platform} & Artix-7 & Artix-7 & Zynq & Artix-7 & ASIC & ASIC$^{\ddagger}$ \\
            \bottomrule
        \end{tabular}
        \begin{tablenotes}
            \footnotesize 
            \item $\dagger$: We select a representative PUF product that has been widely used in the PUF market. It is provided by a company with whom we signed non-disclosure agreements (NDA). We omit its details to respect its intellectual property. More information is provided in \cite{acoregithub} to prove its usability.
            \item $^{\ddagger}$: The frequency of the crystal oscillator in the chip is 30 MHz.
            \end{tablenotes}
    \end{threeparttable}
\end{table}

\begin{figure}[htbp]
    \centering
    \begin{minipage}[c]{0.47\columnwidth}
      \centering
      \includegraphics[width=\textwidth]{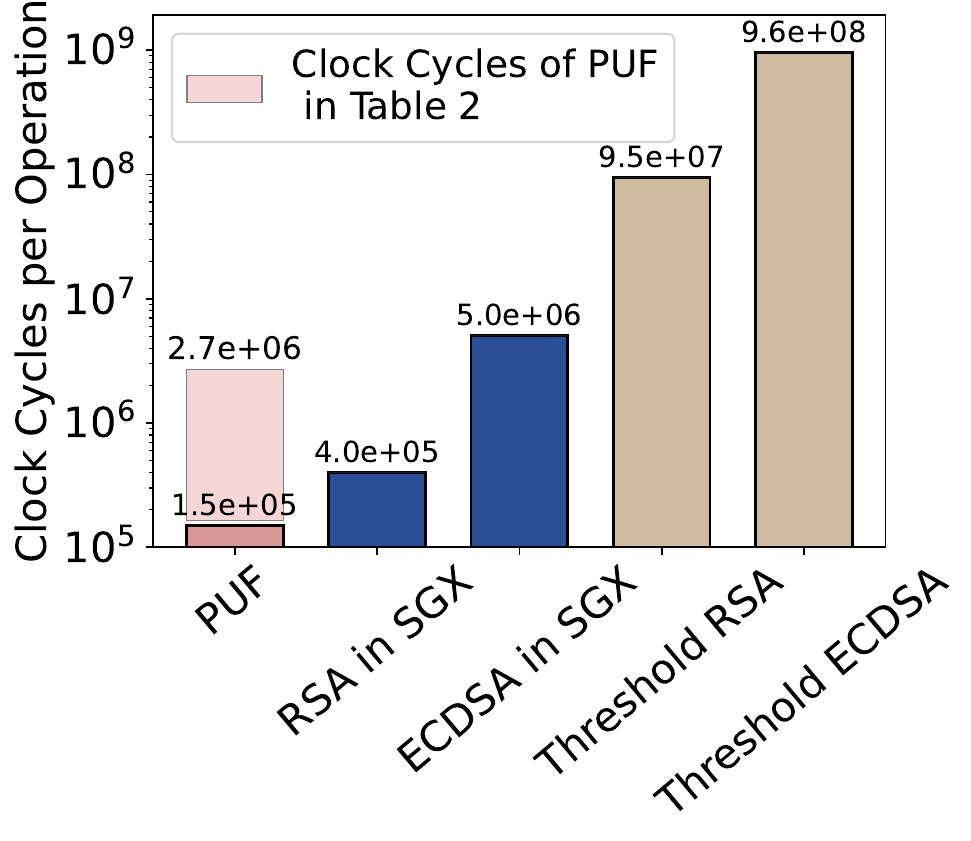}
          \subcaption{Used Clock Cycles}
          \label{fig:puf_cycles_comp}
    \end{minipage}
    \begin{minipage}[c]{0.49\columnwidth}
      \centering
      \includegraphics[width=\textwidth]{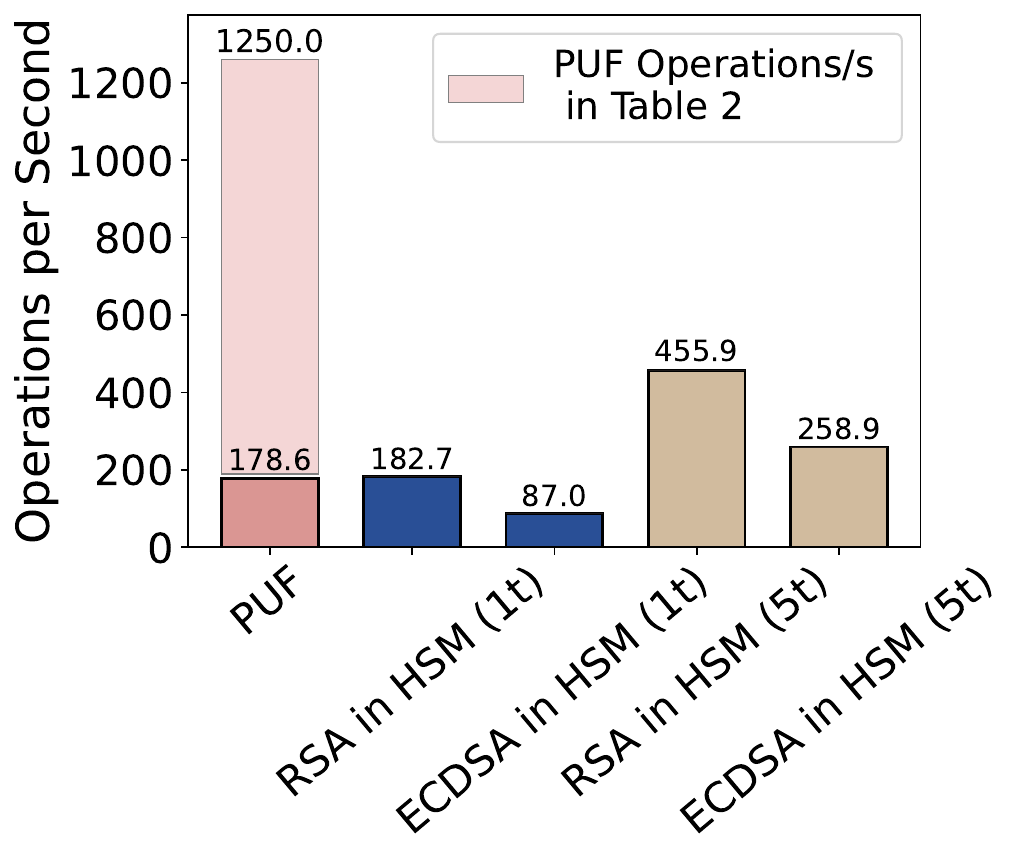}
      \subcaption{Operations per Sec.}
      \label{fig:puf_ops_comp}
    \end{minipage}
    \vspace{-5pt}
    \caption{Performance of CA signing primitives.}
      \label{fig:q3_primitive_comp}
  \end{figure}

We compare PUF with other signing primitives with key protection in Figure \ref{fig:q3_primitive_comp}. We test standard RSA 2048, ECDSA-\texttt{secp256k1} and their threshold signature variants. Since unprotected signing operations are not allowed in the real-world CA software, we execute them in HSM and TEE. Specifically, we choose AWS CloudHSM (one instance) and Intel SGX that both have been used in certificate issuance \cite{cloudhsm}. The crypto implementations are directly ported from OpenSSL-1.1.1t. The HSM signing performance is benchmarked by the official tool \texttt{pkpspeed} while the SGX is tested using \texttt{rdtsc} primitive. As for the threshold variants of RSA and ECDSA, we select two Golang libs\protect\footnotemark that are used in many blockchain applications with $\texttt{signers}=3$ and $\texttt{threshold}=2$.

\footnotetext{See \url{https://github.com/niclabs/tcrsa} and \url{https://github.com/bnb-chain/tss-lib}}

In Figure \ref{fig:puf_cycles_comp}, PUF consumes fewer clock cycles than signing algorithms within SGX. The environment switch caused by TEE can incur 30\%$\sim$40\% overhead. In Figure \ref{fig:puf_ops_comp}, only with the help of multithreading (\eg, 5 threads) can the AWS CloudHSM instance outperform one PUF instance. The operations inside a PUF, such as arbiter delaying, can be quite fast with only several microseconds \cite{PUFarbiterperf2004}. Although  powerful large-scale HSM products like Thales Protect Server can have better performance, our design can also use the faster PUFs with more available ones in the future. Thus, PUF will not be the performance bottleneck in certificate issuance.

\textbf{Comparison with other designs}. We present a comprehensive comparison in Table \ref{tab:design_comparison} in the appendix. Compared to other works, \acore\ has distinct advantages since it is the only one using physically trusted CA endorsements. Also, it does not introduce extra actors and change the original workflow. Due to orthogonal purposes, \acore\ can complement with the listed works such as IKP and F-PKI to achieve full protection for CA.


\noindent\textbf{Answer to Q1:} Compared with traditional techniques, PUF provides stronger security while ensuring the performance. By leveraging its physical bindings, \acore\ sits in a different position from other designs to achieve the secure keyless issuance for CAs.



\subsection{Q2: Microbenchmark}\label{sec:evaluation_microbm}

In this test, we evaluate the overall performance of our functionality prototype. Here we set $M=2$, \ie, two intermediate CAs, and thus form a four-level certificate chain with one root CA and one domain. We set the latency of the simulated PUF to 3 ms (PUF$_{6}$ in Table \ref{tab:puf_evaluation}).

As shown in Figure \ref{fig:q1-ops}, Pebble with \acore\ takes about 200$\sim$350 ms to issue a complete chain. It requires less than 7 ms and 15 ms to perform PIV logging and client verification, respectively. The operations of the CA and logger account for about 93\% of the overall time. The generation of PIV chain takes less than 1.5 ms because it is mainly the lightweight hash calculation. PUF performance accounts for at most $7\%$ in the issuance process, which is relatively small compared to the other operations. 

\begin{figure}[htbp]
    \centering
    \includegraphics[width=0.95\columnwidth]{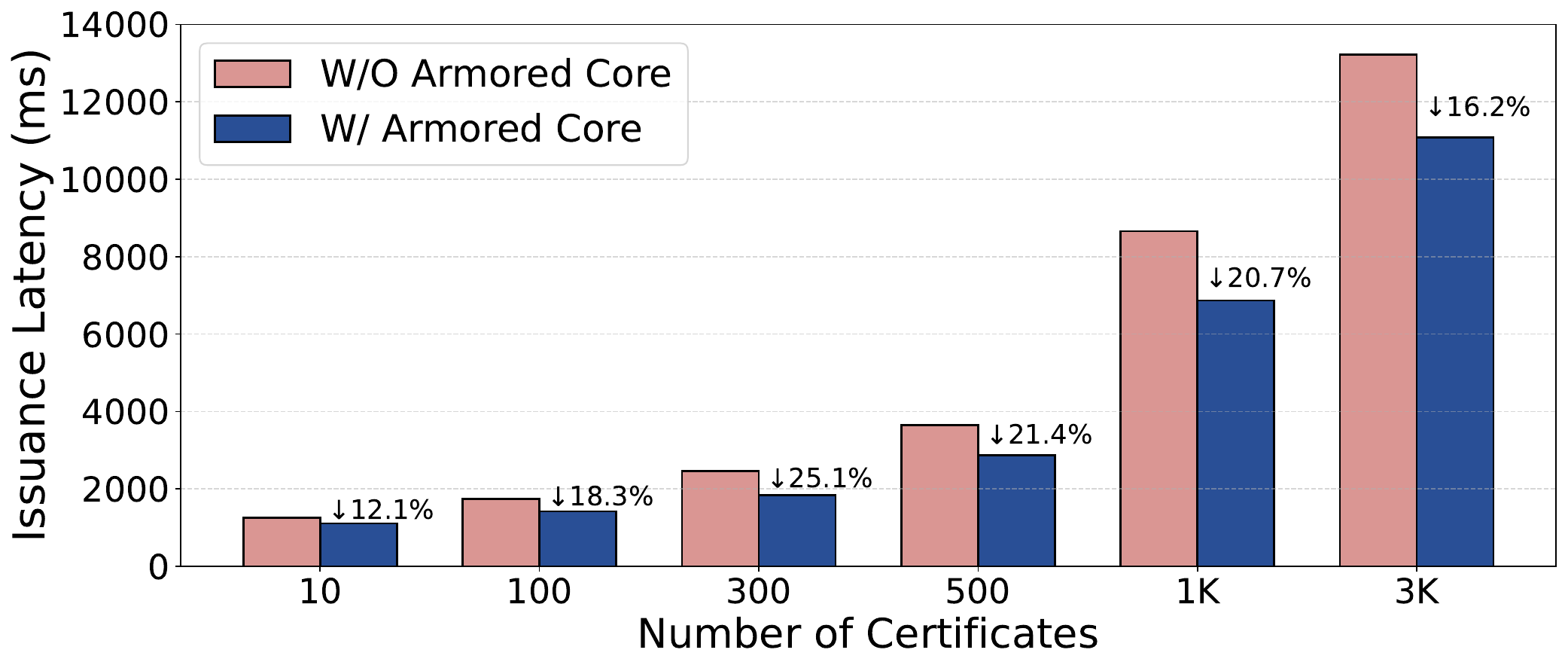}
    \caption{Overhead of certificate issuance.}
    \label{fig:cert_issuance_overhead}
\end{figure}

Moreover, Figure \ref{fig:cert_issuance_overhead} shows the issuance performance as the certificate numbers increases with (W/) and without (W/O) our design. We can see that our PUF-based functions always outperform the original ones. The advantage is gradually reduced because there is a constant overhead for PUF operations. The results can still indicate that \acore\ does not sacrifice the issuance performance for stronger PUF-based security.

\noindent\textbf{Answer to Q2:} The above breakdown results show that the \acore-enabled PKI system is functionally effective and even more efficient than original functions in the issuance process.

    

     


\subsection{Q3: Introduced Overhead}\label{sec:eval_runtime_overhead}

Now we evaluate the detailed overhead that \acore\ can bring to the original PKI system. 

\textbf{Storage.} Table \ref{tab:cert_size_compare} shows that the new certificates are 20\%$\sim$27\% smaller for intermediate CAs. This is because the size of \texttt{sig} and \texttt{pk} fields are aligned with the PUF response length, \ie, 256 bits here. Although the root CA is not necessarily changed, we still need an extra new certificate as a root of PUF-based operations. As for PIVs in a four-level chain, they are encoded to 579 bytes. Therefore, our design in total only requires 2.8\% more storage than the originals. 


\begin{table}[H]
    \centering
    \caption{Size of new certificates for CAs and Domain (bytes).} \label{tab:cert_size_compare}
    \fontsize{8pt}{8pt}\selectfont
    \begin{tabular}{cccccc}
        \toprule
        \multirow{2}{*}{\textbf{Fields}} & \textbf{Root CA} & \multicolumn{2}{c}{\textbf{Intermediate CAs}} & \multicolumn{2}{c}{\textbf{Domain}} \\
        \cmidrule(r){3-4}\cmidrule(r){5-6}
        ~ & (Additional) & before & after & before & after \\
        \midrule
        \texttt{sig} & 32 & 256 & 32 & 256 & 32\\
        \texttt{pk} & 32 & 256 & 32 & - & - \\
        \texttt{ext} & - & - & +32 & - & +32 \\
        Total & \cellcolor{black!10}883 & 2.4K & \cellcolor{black!10}1.9K & 952 & \cellcolor{black!10}688 \\
        \bottomrule
    \end{tabular}
\end{table}

\begin{figure}[htbp]
    \centering
    \includegraphics[width=0.95\columnwidth]{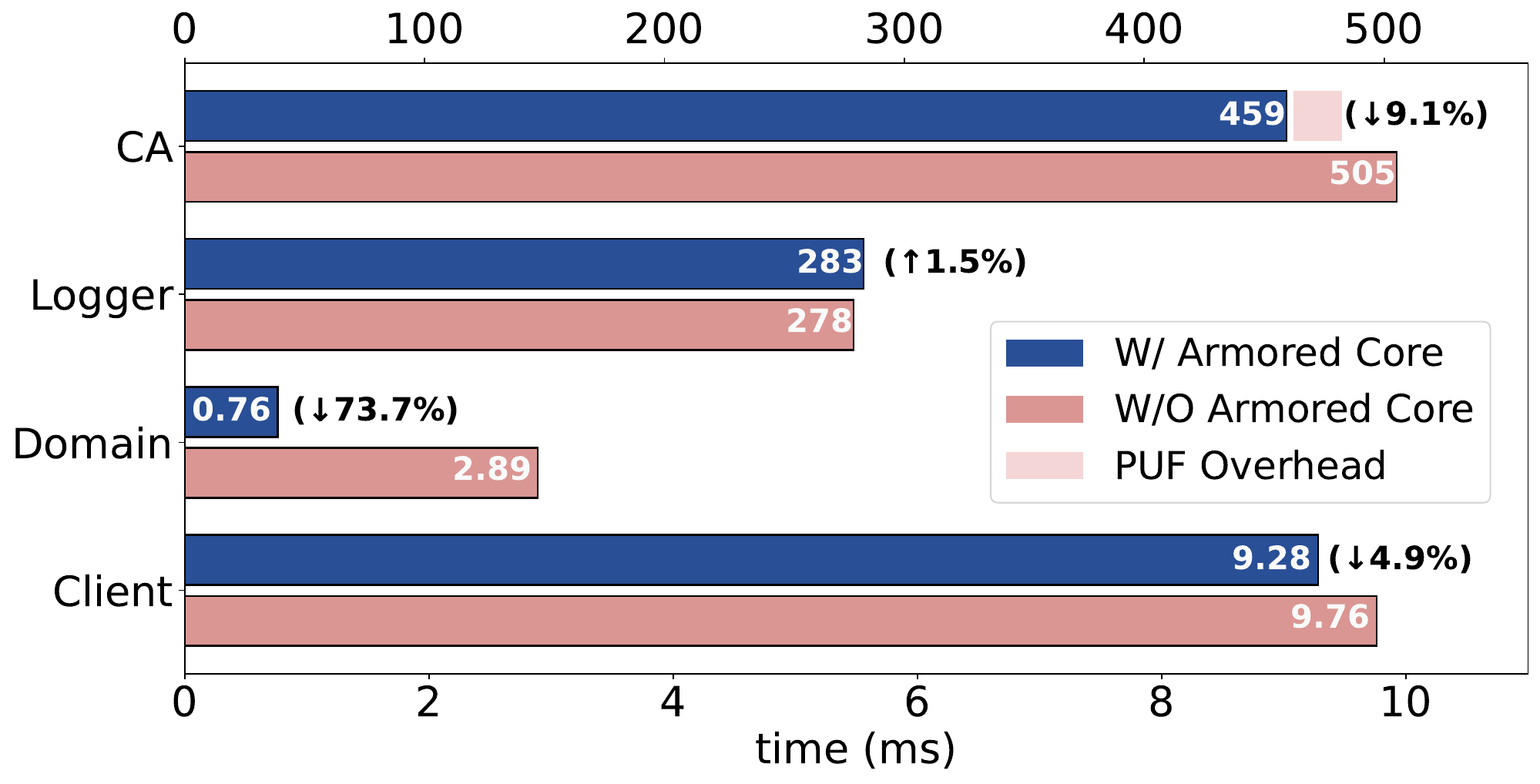}
    \caption{Introduced Overhead on different PKI actors.}\label{fig:q2-overhead}
\end{figure}

\textbf{Computation.} Figure \ref{fig:q2-overhead} gives the end-to-end runtime overhead of \acore. The results show that, most of our functions can outperform the original code, even plus the PUF operations. Our prototype can achieve $4.9\%\!\sim\! 73.7\%$ computational improvement though it is not optimized yet. Only PIV logging only introduces a reasonable delay of average 1.5\% to the Trillian-based logger.

\begin{figure}[htbp]
    \centering
    \includegraphics[width=0.95\columnwidth]{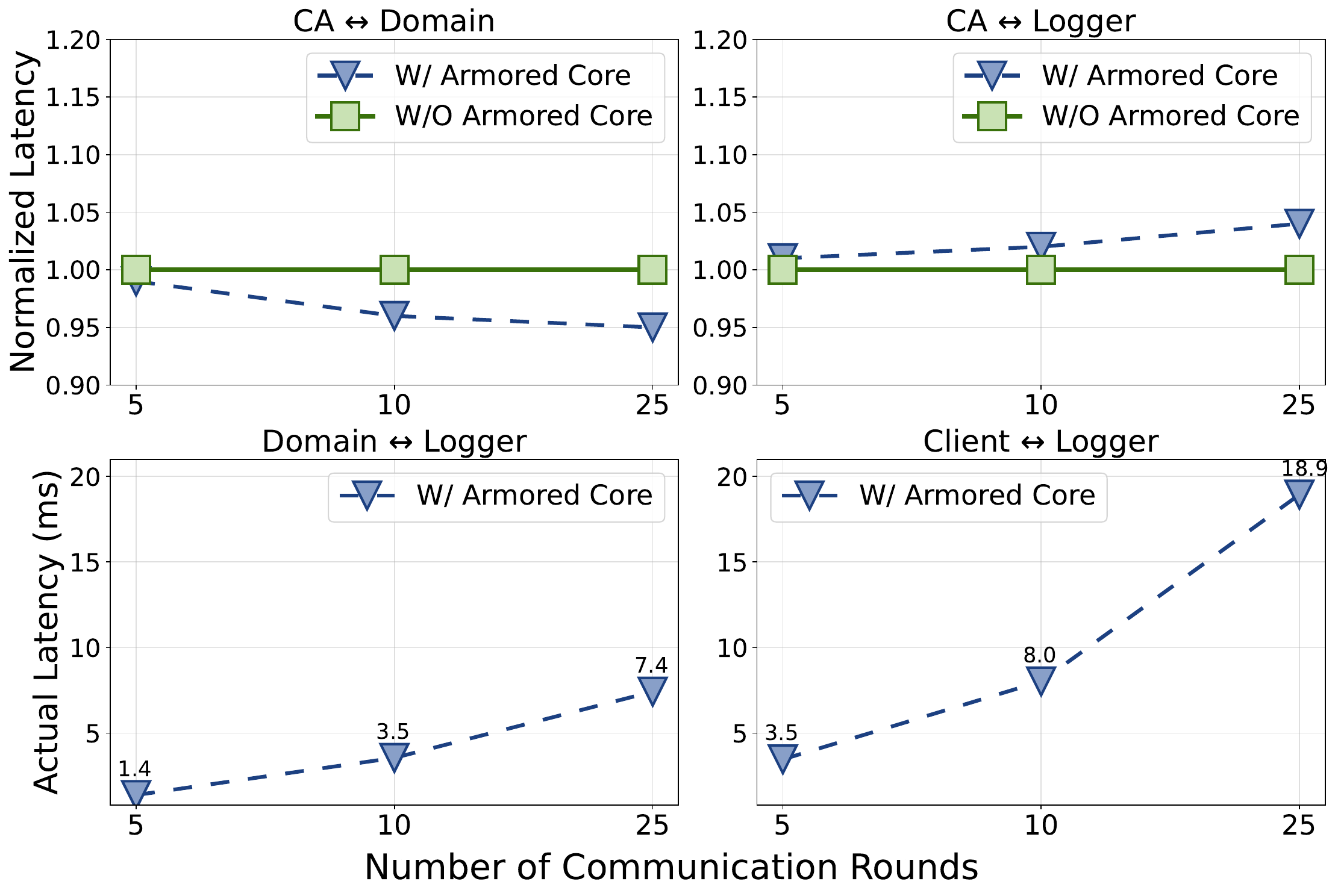}
    \caption{Communication overhead between PKI actors.}
    \label{fig:cert_communication_overhead}
\end{figure}

\textbf{Communication.} We measured the communication latency in our local network as shown in Figure \ref{fig:cert_communication_overhead}. The smaller certificate size slightly reduces the latency by $\sim\! 3.8\%$ between the CA and domain. And the PIVs causes $\sim\! 2.6\%$ increase for the loggers. As for the domains and clients, we record the actual latencies of PIV delivering. They indicate the solid extra overhead that integrating \acore\ may bring to a complex CT system. 


\noindent\textbf{Answer to Q3:} \acore\ can improve the performance of the original system due to the replacement of heavyweight signing operations. The results indicate that it realizes 4.9\%$\sim$73.7\% improvement on computation and only cause $<4\%$ overhead on storage and communication.






\subsection{Q4: Hardware Deployment Cost}\label{sec:eval_deployment_cost}

The introduction of PUF hardware helps \acore\ break the trade-off between security and performance, but it incurs extra cost on equipment updates. Here we evaluate two ways of adopting PUF hardware in CA servers: integration style and standalone style.


\textbf{Integration style.} The average resource utilization of our PUF-embedded FPGA prototype on the board is 13.1\%$\sim$15.3\%, given in Table \ref{tab:acore_hw_performance} in the appendix. Our prototype is resource efficient even with an integrated IPUF instance. 


\textbf{Standalone style.} A more welcomed way is to build a specialized secure element with multiple PUF instances inside. It can be plugged into a server without extra modifications. Unfortunately, there is no such commercial product for now. We can only give a rough estimation of the cost as shown in Table \ref{tab:deployment_cost} since any suitable PUF designs can be adopted here. We can see that PUF has lower power consumption and costs than HSM because it is directly built on electronic units like memory cells. It can significantly decrease energy cost and financial cost for cloud vendors. 



\begin{table}[htbp]
    \setlength{\tabcolsep}{10pt}
    \centering
    \begin{threeparttable}
    \caption{Estimation of hardware deployment cost.} \label{tab:deployment_cost}
        \fontsize{8pt}{8pt}\selectfont
        \begin{tabular}{C{.25\columnwidth}C{.23\columnwidth}C{.23\columnwidth}}
            \toprule
            \textbf{Estimation} & \textbf{PUF}* & \textbf{HSM}* \\
            \midrule
            Power (W) & 30 mW$\sim$1 W & 150 mW$\sim$140 W \\
            Energy (kWh)$\dagger$ & \cellcolor{black!10}0.02$\sim$0.72 & 0.1$\sim$101\\
            \midrule
            Acquisition (USD) & $<$ 10 & 950$\sim$20K \\
            Expense (USD) $\dagger$ & \cellcolor{black!10}$<$ 200 & 19K$\sim$400K \\
            \bottomrule
        \end{tabular}
        \begin{tablenotes}
            \footnotesize 
            \item $\dagger$: We assume that there are 20 installed instances running for one month.
            \item *: For PUF, the statistics is from Table \ref{tab:puf_evaluation} and mainstream PUF companies. For HSM, we checked common products like Yubico HSM, LunaHSM, BlackVault HSM, etc.
        \end{tablenotes}
    \end{threeparttable}
\end{table}








\noindent\textbf{Answer to Q4:} \acore\ is cost-effective and its hardware cost can be afforded by most vendors. It has reasonable financial and energy cost to be deployed in real-world systems.


\section{Discussion and Limitations}\label{sec:discussion}




\noindent$\bullet$ \textbf{Additional benefits of introducing PUF.} We observe that PUFs can bring two promising benefits for PKI other than the keyless issuance, which further supports our design philosophy.

\ding{172} \textit{Reducing the post-compromise damages.} Even if attackers can somehow compromise the whole CA system, they still need that specific CA server because the operations are bound to PUF. Such misissuance is easier to be identified and restricted. In traditional protections, there are no restrictions once attackers recover the key string. Therefore, our design still helps even after the compromise.



\ding{173} \textit{Mitigating the weak key issues.}  PUFs can offer guaranteed performance because it is mandatory \cite{PUFISOP1,PUFISOP2} for manufacturers to pass the required tests on PUF's properties like randomness \cite{turan2018NISTRandom}, reliability, and others. This can mitigate the widely identified weak key issue \cite{heninger2012miningPQ, chi2023detweakkey, hebrok2023TLSsession} in PKI vendors. Even if one instance is defective, the risk is much lower than a weak CA signing key.









\noindent$\bullet$ \textbf{PUF as a Service.} Inspired by HSM-as-a-service in CloudHSM \cite{cloudhsm}, we can further establish PUF-as-a-Service (PaaS) based on the proposed PUF abstraction for server environments. PUFs can be packaged as a remote service so the vendors can rent them for users who require physically trusted proof of their code genuinely running on the expected platforms. CRPs will be the \textit{commitments} for verification. Compared with CloudHSM, PaaS have more lightweight interfaces, and lower manufacturing costs. It has the potential to be a new commercial alternative in the future.






\noindent$\bullet$ \textbf{Limitations and future work.} The advantage of \acore\ actually yields its limitation, \ie, using PUF in the core functions of CA. The hardware modification on servers hinders instant deployment in practice. The aging and hardware failure of PUF requires extra human efforts. For example, the operators may have to change the broken PUF instances in the server. Additionally, the reliability of hardware-generated responses can still be a concern for large-scale use. We argue that some of these limitations can be alleviated with the development of PUF technology.


For the future work, we envision two directions to further explore the use of PUF in PKI. First, we can build a PUF-based root store mechanism on endpoint devices to enhance the protection and mitigate the ``hidden root'' issue \cite{LL0DLZ21hiddenRoot,TraceRootsMa}. Second, by applying the physical binding of PUF into other certificate fields, we can improve the certificate ownership control \cite{Ma21CertName}. 




\section{Related Work}



In the past decade, a substantial body of works have contributed to the improved security design in PKI.


\noindent$\bullet$ \textbf{Enhanced design of PKI/CA}. ARPKI \cite{BasinARPKICKPSS14} is an attack-resilient PKI with an extra CA and a logger against participant corruption. IKP \cite{MatsumotoRIKP17} use blockchain to capture CAs' misbehaviors with automatic detection. CAPS \cite{Matsumoto20CAPSACSAC} has a signaling mechanism to promptly defend against malicious CAs. F-PKI \cite{ChuatKMPFPKI22} introduces a map server to allow clients to verify certificates based on different trust levels defined by domains themselves. DV (Domain Validation) ++ \cite{brandtDomainValidation2018} transforms the centralized validation into a multi-vantage points approach. Duan \etal propose RHINE \cite{RHINEDuanFL0BP23}, an authenticated end-to-end DNS record system. It uses delegation transparency to offer robust authentication and resistance against CA compromises. 

CertLedger \cite{KubilayKM19CertLedger} formalizes a blockchain-based bulletin board model for public transparency. Toorani \etal \cite{toorani2021decentralized} employ PBFT consensus algorithm to confirm the proof of domain certificates. Unlike the previous designs focusing on detecting or resisting the corrupted CAs, \acore\ aims to prevent the CA compromises from key extraction attacks. Therefore, it can naturally fit into those designs with distinct protections.









\noindent$\bullet$ \textbf{PUF-based security designs}. PUF has been used as a security primitive in many scenarios. Chatterjee \etal design a PUF-based anonymous authentication protocols \cite{chatterjeeBuildingPufBased2019, chaterjee3PAARivateUF2021} without the explicit storage of CRPs. PUF-RAKE \cite{qureshiPUFRAKEPUFBasedRobust2022} is a key exchange protocol with CRP obfuscation to establish keys quickly between entities. Zheng \etal \cite{zhengPUFbasedMutualAuthentication2022} propose a provably secure PUF-based authentication protocol for lightweight peer-to-peer communication. SPEAR \cite{xiaolinSPEAR} is a dedicated PUF-based authenticated encryption algorithm. PUF can also be used in remote attestation schemes. Xia \etal \cite{SGXFPGAXiaLX021} design a PUF-based attestation protocol for CPU-FPGA heterogeneous environment. JANUS \cite{zhang2024teamwork} introduces PUF into TEEs to provide intrinsic trust and nested measurements. 

Until now, few studies \cite{PUFSSL2018, PUFPKIIoT2022} have explored leveraging PUF in PKI/CA but they only treat PUF as a key generator (Table \ref{tab:design_comparison}), where the signing key can still be exposed by attacks or operational errors. To our best knowledge, \acore\ is the first to apply the physical binding of PUF for keyless certificate issuance in CAs. 


\noindent$\bullet$ \textbf{Transparency log designs}. CONIKS \cite{CONIKSMelaraBBFF15} utilizes verifiable unpredictable functions to compute the storage index. AAD \cite{TomescuBPPTD19AAD} is an authenticated transparency log based on bilinear accumulators. Merkle$^{2}$ \cite{HuHKYP21Merkle2} is a nested Merkle tree structure for lightweight auditing. TAP \cite{ReijsbergenM0D023TAP} realizes zero-knowledge range proofs for independent audits. The PUF transparency can be built upon these designs.








\section{Conclusion}


In this paper, we propose \acore, a security extension of PKI that achieves keyless certificate issuance using PUF. It can co-exist with traditional signature-based functions and workflows. It enables PUF-based public key endorsement with modified certificate fields. Its PUF transparency logging mechanism ensures the auditability of PUF invocations of CA. We have integrated \acore's functions into real-world PKI software including Pebble and Certbot. The evaluation results show that it is fully compatible with the original workflow to provide enhanced PKI functions without relying on CA's signing key. Also, this integration incurs small overhead on storage and communication, and offers improvements on computation (4.9\%$\sim$73.7\%). In conclusion, \acore\ is an efficient and secure PKI enhancement for security-intensive scenarios. It can establish a new line of research to build a more trusted PKI system on physical trust. 

\begin{table*}[htpb]
  \footnotesize
  \centering
    \begin{threeparttable}
	\caption{Detailed comparison with existing security designs of PKI/CA. \label{tab:design_comparison}}
	\begin{tabular}{rcccccccc}
		\toprule
		\multirow{2}{*}{\textbf{Approach}} & \multirow{2}{*}{\textbf{Motivation}$^{1}$} & \multirow{2}{*}{\textbf{Type}$^{2}$} & \multirow{2}{*}{\textbf{CA Endorsements}} & \multirow{2}{*}{\textbf{Defense Mechanism}} & \multicolumn{4}{c}{\textbf{Introduced Changes}} \\
        & & & & & \textbf{Hardware} & \textbf{Workflow} & \textbf{New Entity} & \textbf{Scale$^{3}$}\\
		\midrule
    \multicolumn{9}{c}{\cellcolor{gray!15}\textsc{Industry}}\\
    Hardware Key Seal & Key exposure & \textit{E} & Signature & HSM, TEE, etc & \compfull & \compnone & \compnone & \compnone \\
    PUF-backed Key Gen. & Key exposure & \textit{E} & Signature & PUF & \compfull & \compnone & \compnone & \compnone \\
    \hline
    \multicolumn{9}{c}{\cellcolor{gray!15}\textsc{Academia}}\\
    ARPKI \cite{BasinARPKICKPSS14} & CA corruption & \textit{E} & Signature & Redundant Checks & \compnone & \comppart & \compfull & \comppart \\
    IKP \cite{MatsumotoRIKP17} & CA corruption & \textit{R} & Signature & Blockchain & \compnone & \compfull & \compfull & \compfull \\
    CertLedger \cite{KubilayKM19CertLedger} & CA corruption & \textit{R} & Signature & Blockchain & \compnone & \compfull & \comppart & \compnone \\
    CAPS \cite{Matsumoto20CAPSACSAC} & CA corruption & \textit{E} & Signature & Signaling set & \compnone & \comppart & \comppart & \compnone \\
    F-PKI \cite{ChuatKMPFPKI22} & CA corruption & \textit{E} & Signature & Customized policy & \compnone & \comppart & \compfull & \compfull \\
    RHINE \cite{RHINEDuanFL0BP23} & DNS-side attacks & \textit{R} & Signature & Trusted delegation & \compnone & \compfull & \compfull & \comppart \\
		\textbf{\acore} & Key exposure & \textit{E} & \textbf{PUF} & \textbf{Physical binding} & \compfull & \compnone & \compnone & \compnone\\
		\bottomrule
	\end{tabular}
    \begin{tablenotes}
	    \small 
        \setlength{\multicolsep}{0cm}
        \begin{multicols}{2}
            \item[1] The threats or attacks against PKI that motivate the design.
            \item[2] \textit{E}: Enhancement of current systems; \textit{R}: Redesign; 
            \item[3] Require globally full-scale deployment to be effective;
            \item[] \compfull\ Changed/Yes; \comppart\ Partially; \compnone\ Unchanged/No; 
        \end{multicols}
    \end{tablenotes}
 \end{threeparttable}
\end{table*}

\begin{table}[htbp]
  \centering
  \caption{Resource utilization of hardware prototype.} \label{tab:acore_hw_performance}
  \fontsize{8pt}{8pt}\selectfont
  \begin{tabular}{ccccc}
      \toprule
      \textbf{FPGA Resources} & \textbf{LUT} & \textbf{FF} & \textbf{BRAMS} & \textbf{DSP} \\
      \midrule
      Synthesize & 61137 & 35670 & 48 & 27 \\
      Percentage (\%) & 29.9 & 8.7 & 10.7 & 3.2 \\
      \hline
      Place and Route & 73337 & 46415 & 49 & 27 \\
      Percentage (\%) & 35.9 & 11.3 & 11.0 & 3.2 \\
      \bottomrule
  \end{tabular}
\end{table}






\begin{acks}

\end{acks}

\bibliographystyle{ACM-Reference-Format}
\bibliography{PUF-PKI.bib}

\appendix

\section{Data Availability}

All software and hardware code developed during this research has been made publicly available at \url{https://anonymous.4open.science/r/Armored-Core-Artifact-4E28/README.md}. The code can be accessed freely. Additionally, we have provided basic instructions to facilitate the deployment of our prototype, enabling other researchers to replicate our experiments and build upon our work.

\section{Security Proof Details}\label{sec:full_proof}

We first formalize the security properties of a keyed cryptographic hash function $H_{k}(\cdot)$.

\begin{definition}[\textbf{AXU Hash}]
	\label{axu-hash-def}
	$H$ is an $\epsilon_{H}$-\textit{almost XOR universal} ($\epsilon_{H}$-\textit{AXU}) hash function, if $\forall M,\ M^{'}\in\bin^{*},\ M \ne M^{'}$ and $\forall c \in \bin^{n}$,
	\begin{equation}
    \small
		\label{axu-hash-prob}
		\mathrm{Pr}\left[H_{k}(M)\oplus H_{k}(M^{'})=c\ \middle|\ k \sample \bin^{n}\right] \le \frac{1}{2^{n}}\cdot\epsilon_{H}.
	\end{equation}
	
	The AXU definition is another way of expressing the collision resistance of $H$ when $c=0$. Thus, an AXU hash function is necessarily collision-resistant. 
\end{definition}

We adopt the classic reduction-based proof technique to derive the probability upper bound of $\Aa$ winning the security games. As stated in \cref{sec:security_analysis}, $\Aa$ is a PPT adversary whose operations are bound in polynomial time and size constraints. 

\subsection{Security Proof of Theorem \ref{thm:thm_1}}\label{sec:full_proof_1}

\begin{theorem}\label{thm:thm_1}
	When $H$ is a standard cryptographic hash function with collision resistance and XOR universality, if PUF $\Pp$ is $\epsilon_{P}-$secure PUF family, then for any PPT adversary $\Dd$ that makes at most $q$ signing queries, its advantage $\mathtt{Adv}_{\Dd}^{IND}$ of distinguishing between $\texttt{sig}$ and random strings is negligible.
\end{theorem}

In Theorem \ref{thm:thm_1}, $\Aa$'s goal is to distinguish the output of the certificate issuing oracle $\Oo_{I}$ of \acore\ between the output of a random bit oracle $\$(n)$. It is allowed to make at most $q$ queries to obtain the signed certificates with submitting certificate entries to $\Oo_{I}$. Also, we assume that the length of submitted certificate chains during the queries is at most $l_{M}$.

\begin{proof}
To prove the distinguishing advantage of $\Aa$, we now define two games $G_{1}$ and $G_{2}$, where $G_{1}$ is the actual certificate issuing oracle of \acore. $G_{2}$ is identical to $G_{1}$ except that the invocation of PUF is replaced with the collision-free random sampling of $\$(n)$.

The purpose of $G_{1}$ and $G_{2}$ is to isolate the observation on PUF operations, thereby reducing the overall security of the design to the PUF primitive. When the adversary is interacting with $\Oo_{I}$ or $\$(n)$, its distinguishing advantage $\mathtt{Adv}_{\Aa}^{IND}$ cannot exceed the probability of collisions of PUF responses. That is, if the generated responses do not collide, $\Aa$ cannot distinguish whether it is interacting with a real \acore\ oracle or the random bit oracle.

Therefore, for all $q$ queries, we have,

\begin{eqnarray}
  \mathtt{Adv}_{\Aa}^{IND} &\le& \tbinom{ql_{M}}{2}\cdot\epsilon_{P} \nonumber \\
  &=& \frac{ql_{M}(ql_{M} + 1)}{2^{n+1}}\cdot\epsilon_{P}
\end{eqnarray}

When $\epsilon_{P}$ is negligible, the advantage of $\mathtt{Adv}_{\Aa}^{IND}$ is also negligible.
\end{proof}

\subsection{Security Proof of Theorem \ref{thm:thm_2}}\label{sec:full_proof_2}

\begin{theorem}\label{thm:thm_2}
	If $\Pp$ is $\epsilon_{P}-$secure PUF family, then for any PPT adversary $\Dd$ that makes at most $q$ certificate issuing queries, its advantage $\mathtt{Adv}_{\Dd}^{EUF}$ in EUF-CMA (Existing Unforgeability under Chosen Message Attack) game is negligible.
\end{theorem}

In Theorem \ref{thm:thm_2}, $\Aa$'s goal is to output a valid pair of certificate chain and the signatures after making at most $q$ queries to an issuing oracle.


\begin{proof}
  We deduce the following statement from the contrapositive of the original theorem: If $\Aa$ wins the EUF-CMA game, then there exists a PPT adversary $\Bb$ who can use $\Aa$ as a subroutine to break the assumptions of $\Pp$. Hence, we can analyze the adversarial advantages by constructing $\Bb$ on $\Aa$.

  In this proof, we fix the length of certificate chains in this proof to $l$. For the $i$-th issuing query and $1 \le j \le l$, $\Aa$ generates $M_{i,j}=\{\texttt{crt}_{i}, \texttt{ts}_{i,j}, N_{i,j}\}$ and sends $M_{i,j}$ to the issuing oracle that is simulated by $\Bb$. $\Bb$ calculates $hc_{i,j}$ as in the original design and sends $hc_{i,j}$ to a PUF oracle $\Oo_{P}$. $\Oo_{P}$ returns a PUF response $R_{i,j}$ to $\Bb$. It then locally calculates $\texttt{sig}_{i,j}$ and returns it to $\Aa$ as the signature.

  In each query, $\Aa, \Bb, \Oo_{P}$ would make such interaction $l$ times to construct the complete certificate chain $\Ll_{i}$. During this procedure, $\Bb$ genuinely constructs a simulated interactive environment for $\Aa$ to make it seem to engage with the real algorithm oracle of \acore. After $q$ queries, let $out_{\Aa}$ represent the final output of $\Aa$ in the EUF-CMA game, where $out_{\Aa}$ contains a set of $M_{j}^{*}$ with the corresponding signatures $\texttt{sig}_{j}^{*}$ where $1 \le j \le l$. A verification oracle $\Oo_{V}$ will capture the result and output the flag bit ($1$ for success and $0$ for failure). $\Bb$ will generate $out_{\Bb}$ according to $out_{\Aa}$. Its output determines whether it succeeds by using $\Aa$'s result.
  
  Now we analyze the probability of $\Bb$ failing in this game. $\Bb$ will be considered to fail in the following three cases.
  \begin{itemize}
    \item \textbf{Case (1)}: $\Aa$ fails, \ie, $\Vv$ outputs $0$;
    \item \textbf{Case (2)}: $out_{\Aa}$ is invalid. At least one of the generated $M^{*}$ cause the collision of $H$. That is, $\exists i\in[1, q], j^{'}\in [1,l]$ such that $hc_{i, j^{'}} = hc_{j}^{*}$;  
    \item \textbf{Case (3)}: $\Aa$ succeeds and $\Vv$ outputs $1$, but $out_{\Bb}$ is invalid.
  \end{itemize}

  The probability of $\Bb$'s failure is at most the sum of the probabilities of these three cases:
  \begin{equation}
    \small
    \prob{\Bb\ \mathrm{fails}} \le \prob{\mathrm{Case (1)}} + \prob{\mathrm{Case (2)}} + \prob{\mathrm{Case (3)}},
  \end{equation}
  where $\prob{\mathrm{Case (1)}}=1 - \mathtt{Adv}_{\Aa}^{EUF}$, $\prob{\mathrm{Case (2)}}=ql^{2}\cdot\epsilon_{H}/2^{n}$. For Case (3), it suggests that $\Aa$ does not create a hash collision but manages to generate a valid \texttt{sig} field. Therefore, we have,
  \begin{eqnarray}
    (1-\frac{ql^{2}\epsilon_{P}}{2^{n}})\cdot (1 - \frac{ql^{2}\epsilon_{H}}{2^{n}}) &\le& (1 - \mathtt{Adv}_{\Aa}^{EUF}) + \frac{ql^{2}\epsilon_{H}}{2^{n}} + \nonumber \\ 
    & & \mathtt{Adv}_{\Aa}^{EUF}\cdot\frac{ql^{2}\cdot\epsilon_{H}}{2^{n}}.
  \end{eqnarray}
  Then we can derive the advantage upper bound for $\Aa$,
  \begin{eqnarray}
    \mathtt{Adv}_{\Aa}^{EUF} &\le& \frac{\frac{ql^{2}(\epsilon_{H} + \epsilon_{P})}{2^{n}} + \frac{ql^{2}\epsilon_{H}}{2^{n}} - \frac{q^{2}l^{4}\epsilon_{P}\cdot\epsilon_{H}}{2^{2n}}}{1-\frac{ql^{2}\epsilon_{H}}{2^{n}}} \nonumber \\
    ~&\le&\frac{\frac{l^{2}(\epsilon_{H} + \epsilon_{P})}{2^{n}} + \frac{l^{2}\epsilon_{H}}{2^{n}}}{1-\frac{l^{2}\epsilon_{H}}{2^{n}}} \nonumber \\
    ~&=&\frac{l^{2}(2\epsilon_{H} + \epsilon_{P} )}{2^{n}-l^{2}\cdot\epsilon_{H}}
  \end{eqnarray}
  Therefore, if the consumption holds ($\epsilon_{P}$ and $\epsilon_{H}$ are negligible values), then the adversarial advantage of forging valid certificates and signatures will also be negligible.
\end{proof}

\section{RISC-V CPU Prototype with Built-in PUF}\label{sec:riscv_platform}

\begin{figure}[htbp]
  \centering
  \includegraphics[width=0.8\columnwidth]{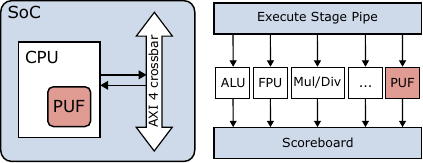}
  \caption{Integration of PUF in RISC-V SoC}
  \label{fig:hw_prototype}
\end{figure}

We provide a preliminary hardware prototype \cite{acoregithub} to prove that current computing infrastructure can satisfy the hardware requirements of \acore. This prototype integrates PUF into the generic CPU on RISC-V platform to offer the built-in PUF functions. We choose CVA6 \cite{RISCVCVA6}, a well-known RISC-V 64-bit CPU project with 6-stage pipeline and single issue, in-order execution. It supports the 64-bit RISC-V instruction set and exhibits good scalability. We can modify or extend some of its components, \eg, adding an FPGA-based PUF module into the CPU, as shown in Figure \ref{fig:hw_prototype}.

The PUF-enhanced CVA6 processor is running on a Genesys 2 FPGA board. As illustrated in Figure \ref{fig:hw_prototype}, the prototype integrates PUF directly into the CPU core, alongside functional units such as the arithmetic logic unit, branch predictor, and multiplier/divider, as a new functional unit in the pipeline execution phase.


Considering the demand for expanding the number of PUFs and the fact that PUF operations may require multiple CPU cycles, we have designed a PUF wrapper \texttt{puf_wrap} to encapsulate PUF instances. The \texttt{puf_wrap} is designed as a finite state machine to control how the module responds to input signals and manages the internal process flow. In our experiment, we instantiated one IPUF \cite{InterposePUF19CHES} within the PUF wrapper, and the \texttt{puf_wrap} itself was instantiated within the execute stage module.

Moreover, we designed a custom instruction, \texttt{pufc} (short for ``PUF challenging"), to utilize the PUF components. This involved modifications to the instruction decode, issue, and commit stages of the CVA6 processor.




\end{document}